\documentclass[twocolumn,amsmath,trackchanges,resetfootnote]{aastex702}

\usepackage[T1]{fontenc}
\usepackage[]{xcolor}
\usepackage{booktabs}
\usepackage{physics}
\usepackage{wasysym}
\usepackage{longtable}
\usepackage{tablefootnote}
\usepackage{threeparttablex} 
\usepackage{tabularx}
\usepackage{graphicx}
\usepackage{textcomp}

\usepackage{array}
\usepackage{amsmath}
\usepackage{changepage}
\usepackage{lineno, ulem} 

\usepackage[flushmargin]{footmisc}
\usepackage{afterpage}
\usepackage{subfigure}
\usepackage{multirow}

\hypersetup{
    unicode=false,                  
    pdftoolbar=true,                
    pdfmenubar=true,                
    pdffitwindow=true,              
    pdfstartview={FitH},            
    pdftitle={OrbParamCatalog},          
    pdfauthor={Nayara},          
    pdfsubject={Astronomy},         
    pdfcreator={dvipdf},            
    pdfproducer={dvipdf},           
    pdfkeywords={}, 
    pdfnewwindow=true,              
    colorlinks=true,                
    linkcolor=red,                  
    citecolor=blue,                 
    filecolor=magenta,              
    urlcolor=violet,                  
    breaklinks=true,
    linktocpage
}

\DeclareUnicodeCharacter{FF0C}{,}

\newcommand{\kms}{\mbox{$\>{\rm km\, s^{-1}}$}}
\usepackage{lipsum}
\usepackage{float}
\usepackage[misc]{ifsym}
\usepackage{placeins}

\shorttitle{Orbital Parameters for 50 Million Stars}
\shortauthors{Feliciano-Souza et al.}

\begin{document}

\title{Orbital parameters for 50 million stars observed by \textit{Gaia} and spectroscopic surveys}

\correspondingauthor{Nayara I. Feliciano-Souza \\ (nayara.nifs@gmail.com)}

\author[orcid=0009-0001-3601-0572]{Nayara I. Feliciano-Souza}
\affil{Departamento de Física (DFI), Universidade Federal de Sergipe, Av. Marcelo Deda Chagas, S/N, 49107-230, São Cristóvão, SE, Brazil}
\email{nayara.nifs@gmail.com}

\author[orcid=0000-0002-0537-4146]{H\'elio D. Perottoni}
\affil{Observat\'{o}rio Nacional, MCTI, Rua Gal. Jos\'{e} Cristino 77, Rio de Janeiro, 20921-400, RJ, Brazil}
\email{hperottoni@on.br}

\author[orcid=0000-0002-7883-5425]{Diogo Souto}
\affil{Departamento de Física (DFI), Universidade Federal de Sergipe, Av. Marcelo Deda Chagas, S/N, 49107-230, São Cristóvão, SE, Brazil}
\email{diogosouto@academico.ufs.br}

\author[orcid=0000-0003-4524-9363]{Friedrich Anders}
\affiliation{Departament de Física Quàntica i Astrofísica (FQA), Universitat de Barcelona (UB), C Martí i Franquès, 1, 08028 Barcelona, Spain}
\affiliation{Institut de Ci\`encies del Cosmos, Universitat de Barceloana (ICCUB), C Mart\'i i Franqu\`es 1, 08028 Barcelona, Spain}
\affiliation{Institut d’Estudis Espacials de Catalunya (IEEC), Edifici RDIT, Campus UPC, 08860 Castelldefels (Barcelona), Spain}
\email{fanders@fqa.ub.edu}

\author[orcid=0000-0002-7662-5475]{Jo\~ao A. S. Amarante}
\affil{Department of Astronomy, School of Physics and Astronomy, \\ Shanghai Jiao Tong University, 800 Dongchuan Road, Shanghai, 200240, China}
\affil{State Key Laboratory of Dark Matter Physics, School of Physics and Astronomy，\\ Shanghai Jiao Tong University, Shanghai, 200240, China}
\email{joaoant@gmail.com}

\author[orcid=0000-0002-9269-8287]{Guilherme~Limberg}
\affiliation{Kavli Institute for Cosmological Physics, University of Chicago, 5640 S Ellis Avenue, Chicago, IL 60637, USA}
\affiliation{Department of Astronomy \& Astrophysics, University of Chicago, 5640 S Ellis Avenue, Chicago, IL 60637, USA}
\email{limberg@uchicago.edu}    

\author[orcid=0009-0001-1380-9488]{Chlo\'{e} Padois}
\affiliation{Departament de Física Quàntica i Astrofísica (FQA), Universitat de Barcelona (UB), C Martí i Franquès, 1, 08028 Barcelona, Spain}
\affiliation{Institut de Ci\`encies del Cosmos, Universitat de Barceloana (ICCUB), C Mart\'i i Franqu\`es 1, 08028 Barcelona, Spain}
\affiliation{Institut d’Estudis Espacials de Catalunya (IEEC), Edifici RDIT, Campus UPC, 08860 Castelldefels (Barcelona), Spain}
\email{chloe.padois@fqa.ub.edu}

\author[orcid=0000-0002-7529-1442]{Rafael M. Santucci}
\affiliation{Universidade Federal de Goi\'as, Instituto de Estudos Socioambientais, Planet\'ario, Goi\^ania, GO 74055-140, Brazil}
    \affiliation{Universidade Federal de Goi\'as, Campus Samambaia, Instituto de F\'isica, Goi\^ania, GO 74001-970, Brazil}
\email{rafaelsantucci@ufg.br}

\author[orcid=0000-0003-2595-5148]{Teresa Antoja}
\affiliation{Departament de Física Quàntica i Astrofísica (FQA), Universitat de Barcelona (UB), C Martí i Franquès, 1, 08028 Barcelona, Spain}
\affiliation{Institut de Ci\`encies del Cosmos, Universitat de Barceloana (ICCUB), C Mart\'i i Franqu\`es 1, 08028 Barcelona, Spain}
\affiliation{Institut d’Estudis Espacials de Catalunya (IEEC), Edifici RDIT, Campus UPC, 08860 Castelldefels (Barcelona), Spain}
\email{tantoja@fqa.ub.edu}

\author[orcid=0000-0002-5274-4955]{Helio J. Rocha Pinto}
\affiliation{Valongo Observatory, Federal University of Rio de Janeiro, Ladeira do Pedro Antônio 43, Saúde, Rio de Janeiro, Brazil}
\email{helio@ov.ufrj.br}

\author[orcid=0000-0002-5084-168X]{Eder Martioli}
\affil{Laborat\'{o}rio Nacional de Astrof\'{i}sica, Rua Estados Unidos 154, 37504-364, Itajub\'{a} - MG, Brazil}
\email{emartioli@lna.br}

\author[orcid=0000-0002-8913-0690]{Arman Khalatyan}
\affil{Leibniz-Institut f\"ur Astrophysik Potsdam (AIP), An der Sternwarte 16, 14482 Potsdam, Germany}
\email{akhalatyan@aip.de}

\begin{abstract}

Together with ground-based spectroscopic stellar surveys, the \textit{Gaia} mission has changed Galactic archaeology over the past decade. Nevertheless, attempts to combine these datasets remain scarce, even though precise orbital parameters for as many stars as possible are crucial to unraveling the Galaxy's formation and evolutionary history. This work aims to construct the largest homogeneous catalog of stellar orbital parameters ever compiled to date, using the most up-to-date data from nine large-scale spectroscopic stellar surveys (\textit{Gaia} RVS DR3, APOGEE DR19, BOSS DR19, LAMOST DR11, \textit{Gaia}-ESO DR5, GALAH DR4, RAVE DR6, SEGUE DR12, DESI DR1), combined with \textit{Gaia} DR3 astrometric data and employing an orbital integration tool for orbital modeling. Our catalog includes orbital parameters, such as orbital distances, eccentricity, inclination, energy, angular momentum, and actions, for 52,423,233 stellar sources from \citealt{BailerJones} distances, and 40,355,569 stellar sources from recomputed \texttt{StarHorse} distances, of which 41,100,866 and 30,777,133 are unique objects, respectively. We also included probabilities of belonging to specific Galactic components based on kinematic classification and chemical-abundance classification. Using this catalog, we recover the known kinematic substructures of the Galactic disk while extending their mapping to Galactocentric distances of $\sim17$~kpc, demonstrating the potential of the combined dataset for studies of Galactic dynamics and evolution. We expect that this catalog will benefit a broad range of research areas, from stellar astrophysics to Galactic archaeology.

\end{abstract}



\section{Introduction}

In the past decades, stellar astrophysics has entered a transformative era driven by large-scale spectroscopic and astrometric surveys that have expanded our observational capabilities. Projects such as APOGEE (Apache Point Observatory Galactic Evolution Experiment; \citealt{APOGEE}), GALAH (Galactic Archaeology with HERMES; \citealt{deSilva2015galah}), RAVE (RAdial Velocity Experiment; \citealt{RAVE}), \textit{Gaia}-ESO \citep{GaiaESO}, SEGUE (Sloan Extension for Galactic Understanding and Exploration; \citealt{SEGUE1}) and LAMOST (Large Sky Area Multi-Object Fiber Spectroscopic Telescope; \citealt{LAMOST}) have collected high-quality spectroscopic data for millions of stars across different regions of the Milky Way. In parallel, the \textit{Gaia} mission \citep{GaiaMission} has revolutionized astrometry by providing precise positions, parallaxes, and proper motions for over a billion stars, and radial velocities for over than 30 million stars.

The \textit{Gaia} mission was designed to precisely map the three-dimensional positions and proper motions of stars throughout the Milky Way and the Local Group \citep{GaiaMission}. Combining astrometry with photometric and spectroscopic capabilities, \textit{Gaia} addresses fundamental questions across astrophysics, from Solar System dynamics to the cosmic distance scale, while its primary goal remains to understand the formation and evolution of the Milky Way \citep{GAIA, GaiaMission}. With a particular focus on the structure, assembly, and dynamical history of the Galaxy, \textit{Gaia} provides an unprecedented stellar census that is transforming our understanding of the Milky Way \citep{Brown2021, Review_LMC2023, Review_starcluster2024, Review_streams2025, Review_satellites2025, Gaia_review}. Together, these surveys enable studies of our Galaxy's structure, formation, and evolution, marking the advent of big data astronomy in stellar research. The combination of precise kinematic and chemical information has opened new avenues for Galactic archaeology, stellar population studies, and dynamical modeling at scales that were previously inconceivable \citep{Review_galacticarchaeology2020, GaiaDR3, Review_galacticarchaeology2024, Hunt2025, Kawata2026}.

However, most existing studies remain limited either to individual surveys or to relatively small, carefully selected samples, and therefore do not fully exploit the combined statistical power of current datasets. This limitation is primarily driven by the heterogeneous nature of the surveys, which differ in instrumental configurations, spectral resolutions, wavelength coverage, and analysis pipelines, leading to systematic differences in derived quantities such as radial velocities and stellar parameters (e.g., \citealt{Tsantaki2022, Turchi2025}).

While combining astrometric and radial velocities gives access to the six-dimensional phase space of stars, these quantities alone do not fully capture their motion. Orbital parameters, by contrast, describe stellar orbits and encapsulate key dynamical aspects. As a result, deriving orbital parameters from observed phase-space data has become a key step in Galactic studies (e.g., \citealt{Trick2019, Hunt2019}). 

In this context, the availability of large samples with homogeneously derived orbital parameters is essential to exploit the current datasets, improving the characterization of known stellar populations, enhancing the detectability of low-contrast structures, and enabling the identification of new dynamical features in the Milky Way, among other applications. Achieving this, however, requires the construction of a large, homogeneous catalog of orbital parameters.

In this work, we address this challenge by combining and homogenizing radial velocities from the latest public data releases of the largest spectroscopic stellar surveys, in order to compute the orbital parameters. The paper is organized as follows: in Section \ref{sec:obs}, we describe all the observational surveys used in this work. In Section \ref{sec:distance}, we describe the adopted distance estimates and compare the impact of different distance catalogs on the derived orbital parameters. In particular, we apply necessary corrections to systematic biases in our reference dataset, the astrometry and spectroscopy from \textit{Gaia}'s third data release (DR3; Section \ref{subsec:gaia}) in Section \ref{subsec:corrections}, and homogenize the radial velocities of the different surveys in Section \ref{sec:rv_norm}. We then use \texttt{AGAMA} \citep{Vasiliev2019} to compute a homogeneous set of orbital quantities for more than 50 million objects with 6D phase-space information, including orbital distances, eccentricity, inclination, energy, angular momentum, and actions (Section \ref{sec:orb_par}). Our final catalog, including our recommendations for selecting high-quality subsamples, is described in Section \ref{sec:catalog}. As an illustrative example of the power of our catalog, we performed a probabilistic dynamical classification into broad Galactic components (thin disk, thick disk, and halo) based on the \cite{Bensby2003} methodology in Section \ref{sec:components}. Our results are presented in Section \ref{sec:results}, while our conclusions are shown in Section \ref{sec:conclusion}.


\section{Surveys}
\label{sec:obs}

In this section, we briefly describe the surveys used in this work, including their main characteristics and the cross-matching procedure (Section \ref{subsec:crossmatch}). A summary of their main properties is presented in Table \ref{tab:survey}.

\subsection{\textit{Gaia} DR3}\label{subsec:gaia}

The current Third \textit{Gaia} Data Release\footnote{\url{https://gea.esac.esa.int/archive/}} provides astrometry and broad-band photometry for 1.8 billion sources and radial velocities for approximately 33 million stars \citep{GaiaEDR3, GaiaDR3}. The astrometric instrument measures positions, parallaxes, and proper motions in the broad optical $G$-band ($330$--$1050$~nm), achieving microarcsecond-level precision for stars brighter than $G\sim15$~mag and extending to $G\sim21$~mag for all sources \citep{GaiaDR3_astrometry}. Complementing this, the Blue Photometer (BP) and Red Photometer (RP) deliver low-resolution spectrophotometry across $330$--$680$~nm and $640$--$1050$~nm, respectively \citep{GaiaDR3_bprpspecvalidation, GaiaDR3_bprpspec}.
The Radial Velocity Spectrometer (RVS; \citealt{GaiaDR2_RVS}) acquires medium-resolution ($R\sim11,000$) spectra centered on the Ca~II triplet ($847$--$874$~nm), currently providing radial velocities for 33 million stars brighter than $G_{\rm RVS}\sim14$~mag \citep{Katz2023}, with typical uncertainties of $\sim1.3$ km~s$^{-1}$ at $G_{\rm RVS}$ $=$ 12, degrading for fainter stars ($>6$ km~s$^{-1}$ at $G_{\rm RVS}\sim14$; \citealt{Katz2023}).

To obtain a unique astrometric reference, we adopt positions and proper motions from \textit{Gaia} DR3, providing a homogeneous and precise foundation for constructing our orbital parameter catalog. Moreover, since \textit{Gaia} provides the largest catalog of positions and radial velocities to date, we also use it to scale the measurements from different surveys (see Section \ref{sec:rv_norm}). Our \textit{Gaia}-based sample comprises 33,652,049 objects with complete positions and radial velocities.

\subsection{APOGEE DR19}
\label{subsec:apogee}

APOGEE is a large-scale near-infrared spectroscopic survey that is part of the Sloan Digital Sky Survey (SDSS; \citealt{SDSS}).
APOGEE is a multi-object high-resolution spectrograph ($R \sim 22,500$) equipped with 300 fibers for simultaneous infrared observations in the H-band 
\citep{Wilson2019}. The survey operates on the Apache Point Observatory (APO) in the Northern Hemisphere, using the Sloan Foundation 2.5-meter and a 1-meter Telescope to observe bright stars \citep{Gunn2006}, and on the 2.5-m Irénée du Pont 2.5-meter Telescope \citep{Bowen&Vaughan1973} at Las Campanas Observatory (LCO) in Chile for Southern Hemisphere observations.

The high spectral resolution of APOGEE enables a typical radial velocity precision of approximately 0.07 km s$^{-1}$ for stars with a signal-to-noise ratio (SNR) greater than 20 \citep{APOGEE}, and delivers detailed abundances for up to 20 elements tracing different nucleosynthetic pathways \citep{GarciaPerez2016}.

In this paper, we use the most recent APOGEE data from the SDSS Nineteenth Data Release (DR19; \citealt{APOGEEDR19})\footnote{\url{https://data.sdss.org/sas/dr19/spectro/astra/0.6.0/summary}}, which is the second release of the fifth phase of the SDSS (SDSS-V; \citealt{SDSSV}). As with previous releases, DR19 is cumulative and includes all APOGEE observations obtained since 2011. This release contains radial velocities available for 1,095,480 stellar sources. 
From this release, we select a sample of 1,076,096 objects with astrometry from \textit{Gaia}.

\subsection{BOSS DR19}
\label{subsec:boss}

The SDSS-V Milky Way Mapper provides optical spectroscopy obtained with the Baryon Oscillation Spectroscopic Survey (BOSS; \citealt{BOSS&SEGUE_spec}) spectrographs, located on the 2.5-m Sloan Telescope at APO and on the 2.5-m du Pont Telescope at LCO. The BOSS spectrographs provide wavelength coverage from approximately 360 to 1000 nm at a spectral resolution of $R\sim2,000$. BOSS can achieve precision of $\sim 4-5 \kms$ for spectra of low-mass stars with high SNR \citep{Kounkel2023}.

In this work, we use the public BOSS data release\footnote{\url{https://dr19.sdss.org/sas/dr19/spectro/boss/redux/v6_1_3/}} included in the SDSS Data Release 19 (DR19; \citealt{APOGEEDR19}), which presents the first release from the repurposed BOSS spectrograph. From this dataset, we select a sample of 822,290 stars, containing only objects with available radial velocity and \textit{Gaia} astrometry.

The survey operates from both hemispheres, with APOGEE-N located at Apache Point Observatory in the northern hemisphere using the Sloan Foundation 2.5-meter and a 1-meter Telescope to observe bright stars \citep{Gunn2006}, and APOGEE-S situated at Las Campanas Observatory in the southern hemisphere, utilizing the Irénée du Pont 2.5-meter Telescope \citep{Bowen&Vaughan1973}.

\subsection{SEGUE DR12}
\label{subsec:segue}

We also incorporate data from the Sloan Extension for Galactic Understanding and Exploration (SEGUE), which was part of SDSS-II and the first year of SDSS-III \citep{SDSS-III}, and consisted of two distinct phases: SEGUE-1 and SEGUE-2 \citep{SEGUE1, SEGUE2}.
SEGUE utilized the SDSS multi-object spectrographs \citep{BOSS&SEGUE_spec} mounted on the 2.5-meter Sloan Foundation Telescope at Apache Point Observatory \citep{Gunn2006}. The system employed 640 optical fibers per plate, enabling simultaneous observations of hundreds of stars. Both SEGUE phases covered the wavelength range 390–920~nm at $R \sim 1{,}800$, reaching targets as faint as $g \sim 20$ \citep{SEGUE1}.  This configuration delivers radial velocities with $\sim 4$--$10$~\kms~precision \citep{SEGUE1}.

For this work, we utilize stellar parameters and radial velocities from the publicly available SEGUE DR12 catalog\footnote{\url{https://data.sdss.org/sas/dr12/sdss/sspp/}}. This dataset provides low-resolution spectroscopy and radial velocities for 377,544 stars, with derived parameters obtained through the SEGUE Stellar Parameter Pipeline (SSPP; \citealt{SEGUE_pipeline}). After cross-matching with \textit{Gaia} DR3 astrometry, we obtain a sample of 368,600 objects.

\subsection{GALAH DR4}
\label{subsec:galah}

The GALAH survey \citep{GALAH,GALAH_Martell} employs the High Efficiency and Resolution Multi-Element Spectrograph (HERMES; \citealt{HERMES_firstlight}) installed on the 3.9-meter Anglo-Australian Telescope (AAT) at Siding Spring Observatory. HERMES is a fiber-fed, high-resolution spectrograph ($R \sim 28{,}000$) capable of simultaneously observing 400 stars across four non-contiguous optical bands spanning approximately 471--789~nm \citep{HERMES_firstlight}. This configuration enables precise radial velocity measurements (0.23 km s$^{-1}$) and detailed chemical abundances for up to 30 elements \citep{GALAHDR4}, supporting investigations of stellar populations, chemodynamical processes, and the formation history of the Milky Way.

For this work, we employ stellar parameters and radial velocities from the fourth data release of the GALAH survey (GALAH DR4; \citealt{GALAHDR4})\footnote{\url{https://cloud.datacentral.org.au/teamdata/GALAH/public/GALAH_DR4/catalogs/}}, which provides high-resolution optical spectroscopy and radial velocity for 1,077,816 stars. From this release, we obtain a sample of 1,062,612 objects with \textit{Gaia} astrometry.

\subsection{RAVE DR6}
\label{subsec:rave}
The Radial Velocity Experiment (RAVE; \citealt{RAVE}) was a medium-resolution spectroscopic survey conducted between 2003 and 2013. 
RAVE employed the Six-degree Field (6dF) multi-object spectrograph \citep{6dF} mounted on the UK Schmidt Telescope at the Anglo-Australian Observatory. The survey obtained medium-resolution spectra ($R \sim 7,500$) centered on the Ca\,\textsc{ii} triplet region (841--879~nm). This spectral range enables radial velocity measurements and estimates of basic stellar parameters such as temperature, gravity, and metallicity. With a typical SNR of 40, RAVE achieves radial velocity precision better than 2~km~s$^{-1}$ for most targets \citep[][see their Figure 8]{RAVEDR6}.

In this study, we use the sixth and final data release of the RAVE survey (RAVE DR6;
\citealt{RAVEDR6, Steinmetz2020})\footnote{\url{https://www.rave-survey.org/metadata/ravedr6}}, from which we adopt the radial velocity measurements. For these stars, stellar parameters and abundances are taken from the MADERA stellar parameters catalog. We crossmatch this sample of 518,387 stars with available radial velocities with \textit{Gaia} DR3 astrometry. Our sample comprises 444,966 stars with phase-space information derived from this cross-match.

\subsection{LAMOST DR11}
\label{subsec:lamost}

LAMOST (\citealt{LAMOST}) is a spectroscopic facility that conducts the large-scale, low- to medium-resolution LAMOST survey. Over the past decade, the survey has collected millions of spectra through two main programs, the LAMOST ExtraGAlactic Survey (LEGAS) and the LAMOST Experiment for Galactic Understanding and Exploration (LEGUE; \citealt{LEGUE, LAMOST_survey}). 

The LAMOST survey, conducted at Xinglong Observatory, employs a 4-meter Schmidt telescope equipped with 4000 fibers distributed across a 5-degree field of view, feeding light into 16 spectrographs connected to 32 CCD cameras \citep{LAMOST}. During its first phase (LAMOST-I, 2012--2017), the survey operated solely in low-resolution mode ($R \sim 1,800$), covering a broad optical wavelength range from 370 to 900~nm \citep{LAMOST_survey}. In 2017, all spectrographs were upgraded to also support a medium-resolution mode ($R \sim 7,500$), marking the beginning of the second phase (LAMOST-II, 2018--2023). This phase includes both the Low-Resolution Survey (LRS),
and the Medium-Resolution Survey (MRS; \citealt{LAMOST_MRS}). 
The MRS operates with a more limited spectral coverage than the LRS, with the blue and red arms spanning 495--535~nm and 630--680~nm, respectively. Nonetheless, the increased resolution enables a higher radial velocity precision of $\sim$1~km~s$^{-1}$ for late-type stars, compared to $\sim$5~km~s$^{-1}$ achieved in the low-resolution mode for high SNR spectra \citep{LAMOST_10yrs, LAMOST_MRS_RV}. While the LRS provides stellar atmospheric parameters and chemical abundances \citep{LAMOST_DR1}, the MRS allows the determination of abundances for a similar number of elements but with significantly improved precision, enabling detailed chemical tagging in Galactic studies \citep{LAMOST_MRS}.

For this study, we use the LAMOST DR11 public data\footnote{\url{https://www.lamost.org/dr11/v2.0/catalogue}}, which includes observations from both the LAMOST-I and LAMOST-II phases of the survey. DR11 comprises data from the LRS and the MRS, and contains measurements for 7,896,309 observations with radial velocities from LRS and 2,575,268 stars from MRS. From this dataset, we select a subsample of 7,706,127 LRS and 2,534,402 MRS stars with available astrometry information after cross-matching with \textit{Gaia}.

\subsection{\textit{Gaia}-ESO DR5} 
\label{subsec:gaia-eso}

The \textit{Gaia}-ESO Spectroscopic Survey is an extensive project that obtained spectroscopy across major Galactic components, including the halo, thick and thin disks, and the bulge \citep{GaiaESO_survey,GaiaESO}.
The \textit{Gaia}-ESO Survey is carried out with the multi-object spectrograph Fiber Large Array Multi Element Spectrograph (FLAMES; \citealp{FLAMES}). It employs both the GIRAFFE spectrograph \citep{GIRAFFE}, with $R \sim 20{,}000$, and the Ultraviolet and Visual Echelle Spectrograph (UVES; \citealp{UVES}), with $R \sim 47{,}000\text{--}50{,}000$ \citep{GaiaESO_survey, GaiaESO_spec}. Together, these instruments cover the spectral range from $370$ to $900$ nm. The high spectral resolution achieved by both instruments enables radial velocity precisions of $0.25~\mathrm{km\,s^{-1}}$ with GIRAFFE and $0.32~\mathrm{km\,s^{-1}}$ with UVES.
In this paper, we use the fifth data release (DR5)\footnote{\url{https://www.eso.org/qi/catalog/show/411}}, which includes 111,348 stars with radial velocities, from which we retain 109,868 objects with available \textit{Gaia} astrometry.

\begin{deluxetable*}{lccccccr}

\tablecaption{Summary of the surveys and samples used in this work. 
\label{tab:survey}}
\tabletypesize{\scriptsize}
\tablehead{
\colhead{Survey} & \colhead{Resolution} & \colhead{Spectral range (nm)} & \colhead{Survey's objects$^{a}$} & \colhead{Sample {\it Gaia} DR3$^{b}$} & \colhead{Survey's uniques$^{c}$}  & \colhead{Sample BJ$^d$} & \colhead{Sample SH$^{e}$}  
}
\startdata
APOGEE DR19            &  22,500         &  1,510--1,700 &  1,095,480    &  1,076,096      &  964,710     &  1,074,161     &  846,575      \\
BOSS DR19              &  2,000          &  365--950     &  1,137,702    &  822,290        &  474,959     &  815,485       &  545,180      \\
DESI DR1               &  2,000--5,500   &  360--982     &  6,372,607    &  4,759,633      &  4,535,597   &  4,745,455     &  1,846,897    \\
\textit{Gaia} RVS DR3  &  11,500         &  845--872     &  33,812,183   &   33,652,049    &  33,812,183  & 33,581,727     &  26,660,281   \\ 
\textit{Gaia}-ESO DR5  &  20,000; 50,000 &  370--900     &  111,348      &    109,868      &  109,866     & 109,333        &  53,993       \\
GALAH DR4              &  28,000         &  470--790     &  1,077,816    &    1,062,612    &  909,618     &  1,062,120     &  960,177      \\
LAMOST LRS DR11        &  1,800          &  370--910     &  7,896,309    &     7,706,127   &  5,557,436   & 7,689,717      &  6,603,183    \\
LAMOST MRS DR11        &  7,500          &  495--680     &  2,575,268    &    2,534,402    &  1,184,065   & 2,533,830      &  2,279,215    \\
RAVE DR6               &  7,500          &  841--879     &  518,387      &     444,966     &  444,966     &  444,006       &  400,646      \\
SEGUE DR12             &  1,800          &  390--900     &  377,544      &   368,600       &  331,313     & 367,399        &  159,422      \\ 
Total                  &  ...            &  ...          &  54,974,644   &    52,537,643   &  48,324,710  &  52,423,233    &  40,355,569   \\ \hline
\enddata
\tablecomments{~\tablenotemark{$^{a}$} objects with avaiable radial velocities from each survey.}
\tablecomments{~\tablenotemark{$^{b}$} remaining objects after cross-matching with the astrometry from {\it Gaia} DR3.}
\tablecomments{~\tablenotemark{$^{c}$} objects from each survey with available radial velocities and Gaia DR3 astrometry, retaining only one measurement per object.}
\tablecomments{~\tablenotemark{$^d$} remaining objects after cross-matching with the astrometry from {\it Gaia} DR3 and Bailer-Jones distances.}
\tablecomments{~\tablenotemark{$^{e}$} remaining objects after cross-matching with the astrometry from {\it Gaia} DR3 and 
{\tt StarHorse} distances using stellar parameters from \citet{Khalatyan2024} as additional input.}

\end{deluxetable*}

\subsection{DESI DR1}   
\label{subsec:desi}

The Dark Energy Spectroscopic Instrument (DESI; \citealt{DESI}) is a spectroscopic survey conducted on the 4-meter Mayall Telescope at Kitt Peak National Observatory and has ten identical spectrographs, each consisting of three arms: blue ($B$: 360–580~nm), red ($R$: 576–762~nm), and near-infrared ($Z$: 752–982~nm), achieving spectral resolving powers between $R \sim 2,000$ and $5,500$ depending on wavelength \citep{Koposov2024, DESI}. Despite the low resolution of DESI, according to \citep{Cooper2023}, the mean radial velocity error is lower than 1\kms~for most stars.

Stellar data from DESI can be obtained from two different pipelines: the RVS pipeline (RVSpecfit) and the Stellar Parameters (SP; \citealt{Koposov2024}). In this work, we use the SP pipeline, which employs the FERRE code\footnote{\url{https://github.com/callendeprieto/ferre}} to derive stellar parameters and chemical abundances, including C, Mg, Ca, and Fe.

Here, we use the first Data Release (DR1)\footnote{\url{https://data.desi.lbl.gov/public/dr1/vac/dr1/mws/iron/v1.0/}}, which covers the first 13 months of the DESI main survey. The stellar component of DR1, available from the Milky Way Survey \citep{Cooper2023}, includes a value-added catalog, the MWS VAC, which presents calibrated radial velocities and atmospheric parameters. The sample of 6,372,607 stars in DESI DR1 yields 4,924,259 objects with \textit{Gaia} astrometry after cross-matching.

\subsection{Catalog Crossmatch}
\label{subsec:crossmatch}

To assign proper motions for the stars in our sample, we cross-match all datasets with \textit{Gaia} DR3 using the \texttt{TOPCAT} \citep{TOPCAT} cross-match tool, with a search radius of 3 arcsec. We double-check our cross-match using the \textit{Gaia} Source Identifier provided in the spectroscopic survey when available. 

For a few objects, we found mismatches. In these cases, for those surveys that provide a \textit{Gaia} identifier (ID), the correct match was performed directly using its {\tt source\_id} through the \textit{Gaia} ESA Archive\footnote{\url{https://gea.esac.esa.int/archive/}}. Some surveys, such as \textit{Gaia}-ESO, RAVE, and SEGUE, do not provide a \textit{Gaia} ID, so we obtain the correct match using the internal cross-match provided for the \textit{Gaia} ESA Archive. For BOSS and DESI, the \textit{Gaia} ID is taken from the second data release, which may not match DR3. For those, we use the \texttt{gaiadr3.dr2\_neighbourhood} in the \textit{Gaia} ESA Archive to obtain the DR3 ID, and then obtain the corrected match. The same approach is applied in Section \ref{sec:distance} to obtain the distances from \cite{BailerJones}.


\section{Distances}
\label{sec:distance}

To determine precise orbital parameters, precise distances are as essential as the kinematic input parameters. With the proper motions from \textit{Gaia} DR3 and the precise radial velocities ($\simeq 3$ \kms) from spectroscopic surveys for the majority of stars, the limited distance precision is often the major contributor to the uncertainty of orbital parameters \citep{Coronado2018}. For stars with precise \textit{Gaia} DR3 parallaxes (e.g., $\varpi/\sigma_{\varpi}\gtrsim5$), the distances are very well constrained by astrometry alone (and $\sigma_{\varpi}/\varpi\simeq\sigma_{d}/d$; see \citealt{Luri2018}). These are mostly stars located within a 3 kpc sphere around the Sun. For stars with lower signal-to-noise parallax measurements, more sophisticated methods have to be used to compute distances \citep{Weiler2025}, and the inclusion of spectro-photometric information often greatly improves the distance precision and accuracy \citep[e.g.,][]{Queiroz2023}.

After cross-matching the spectroscopic surveys with {\it Gaia} astrometry, we performed a second cross-match with two independent distance catalogs. We include distances from both catalogs in the final release, allowing users to select the catalog that best suits their scientific application.  First, we cross-matched our sample with the photo-geometric distance catalog based on \textit{Gaia} Early Data Release 3 (EDR3; \citealt{GaiaEDR3}) provided by \cite{BailerJones}, in order to provide homogeneous distance estimates for as many stars in our catalog as possible. The Bailer-Jones (BJ) catalog provides geometric (parallax) distances for 1.47 billion stars and photo-geometric distances for more than 1.35 billion. The photo-geometric distances incorporate \textit{Gaia} photometry (G and G$_{BP}$ -- G$_{RP}$) to better constrain distances, particularly for stars with uncertain parallaxes.

Nevertheless, we caution users that the \cite{BailerJones} distances have been shown to be significantly biased by the distance prior in the low-signal-to-noise regime parallaxes ($\varpi/\sigma_{\varpi}\lesssim3$). This is due to two widespread conceptual errors in the Bayesian parallax inversion framework, as reported in Section 5 of \cite{Weiler2025}.

We therefore also use distances calculated with the latest version of the {\tt StarHorse} (SH) code \citep[e.g.][]{Queiroz2018, Anders2022} which incorporates the corrections identified by \citet{Weiler2025}. The new {\tt StarHorse} distances for {\it Gaia} DR3 (Khalatyan et al., in prep.) supersede the {\it Gaia} EDR3 results presented in \citet{Anders2022} --- they use the stellar parameters from \citet{Khalatyan2024} for stars with DR3 XP spectra \citep{GaiaDR3_bprpspecvalidation}.

For convenience, throughout the remainder of this paper, results based on the {\tt StarHorse} and \citet{BailerJones} distance estimates are denoted as SH and BJ, respectively. All figures, tables, and analyses explicitly indicate which distance set is adopted. We perform a comparison between the orbital parameters derived from the SH and BJ distances in Appendix \ref{appendix_comparison_SH_BJ} to illustrate the impact of the different distance estimates on the computed orbital parameters.


\section{Systematic Corrections in \textit{Gaia} Data}
\label{subsec:corrections}

For determining orbital parameters, we adopt \textit{Gaia} as the reference source for radial velocities, as it provides a much larger and more homogeneous sample than any other survey in our dataset, offering full-sky coverage and enabling a significantly higher cross-match fraction across all surveys. Nevertheless, systematic effects remain, and corrections to the proper motions and radial velocities, along with their associated uncertainties, are required. These corrections are described in detail below.

\begin{figure*}[htb!]
    \centering
    \includegraphics[width=0.49\linewidth]{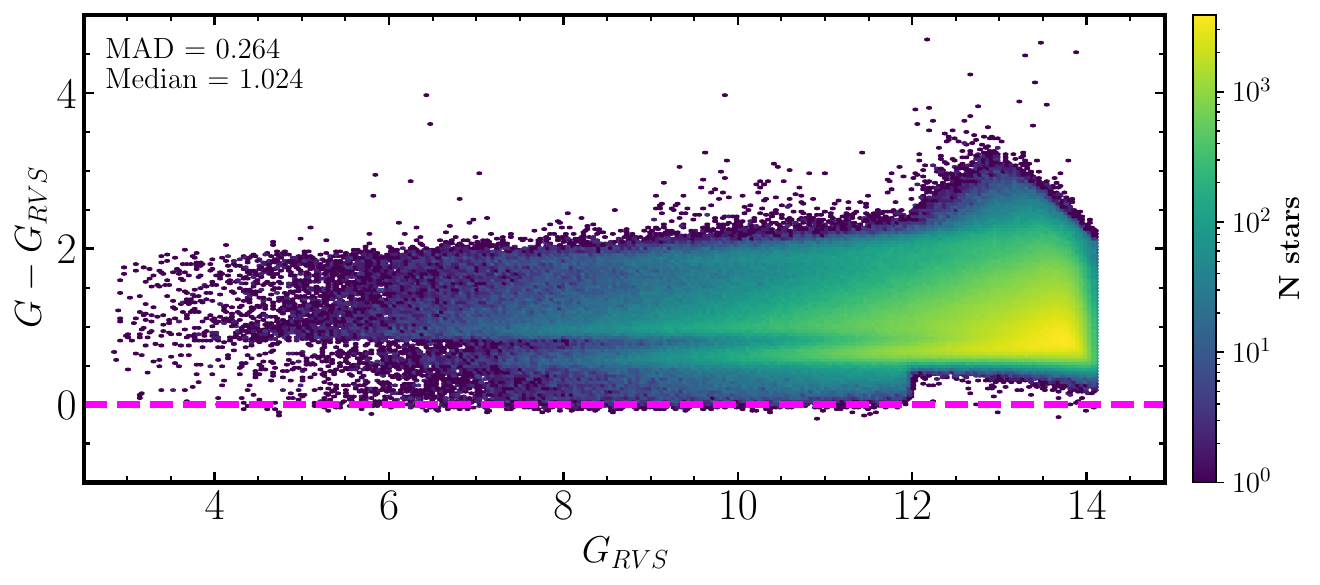}
    \includegraphics[width=0.49\linewidth]{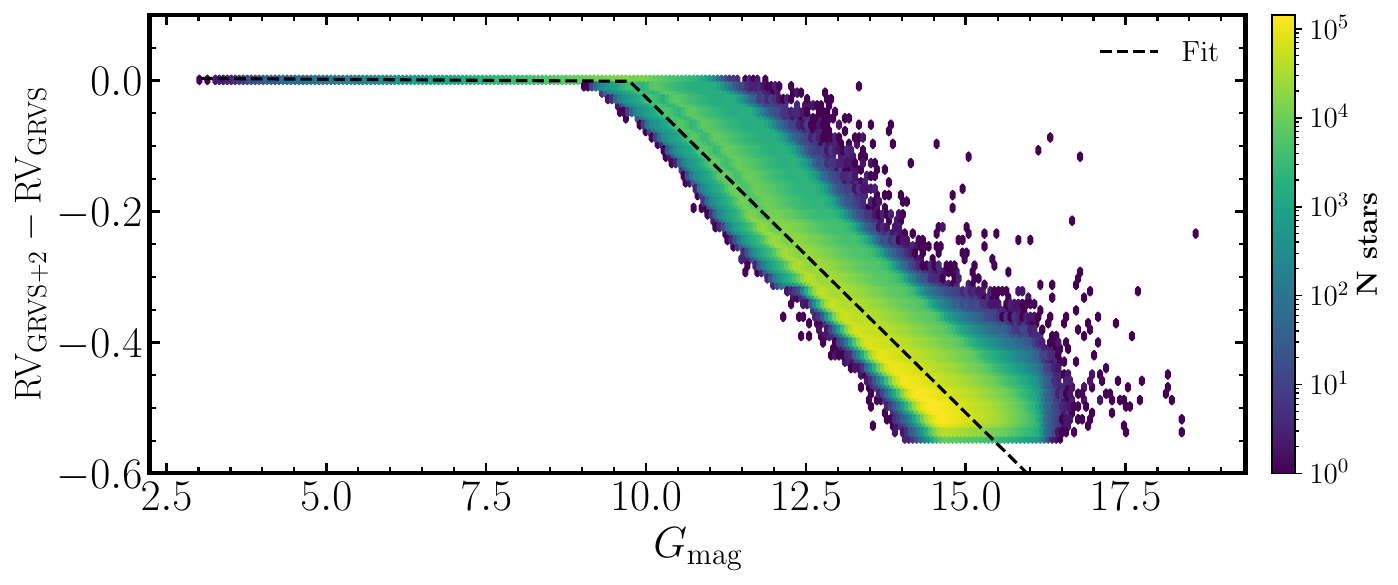}
    \caption{Left panel: Comparison between $G_{\mathrm{RVS}}$ and G magnitude, both from \textit{Gaia}, color-coded by the number of stars. The magenta dashed line indicates the MAD. Right panel: Relation between the G magnitude and the difference in radial velocities obtained using the GRVS + 2 and GRVS corrections for the full Gaia DR3 RVS sample. The black dashed line indicates the best fit for the data, which is $f_{\text{mag}} = 0$ for the straight part, and $f_{\text{mag}} = -0.096314~ G_{\text{mag}} + 0.937386$ for the inclined part.}
    \label{fig:gaia_magnitudes_comparison}
\end{figure*}

We correct the radial velocities following Equation 5 of \cite{Katz2023} for stars with $T_{\mathrm{eff}} < 8,500\,\mathrm{K}$ and Equation 1 of \cite{Blomme2023} for stars with $T_{\mathrm{eff}} \geq 8,500\,\mathrm{K}$, who identified and quantified systematic offsets in the \textit{Gaia}~DR3 radial velocity scale. In addition, we apply the correction proposed by \cite{Babusiaux2023} (see their Equation~1), which introduces a scaling factor to the reported radial velocity uncertainties.

In order to do the corrections described in this Section, we needed the $G_{\mathrm{RVS}}$ magnitude. We note that 1,579,996 stars in \textit{Gaia} RVS do not have a measured $G_{\mathrm{RVS}}$ magnitude. For those objects, we adopt the correction from \cite{Katz2023} using the broad-band $G$ magnitude as a proxy. Although the $G$ and $G_{\mathrm{RVS}}$ magnitudes are generally similar, with a mean absolute deviation (MAD) of $0.264$~mag and a median of $1.024$~mag, they are not identical and can differ by up to $2$~mag in extreme cases (see Figure~\ref{fig:gaia_magnitudes_comparison}, left panel). When applying Equation 5 of \cite{Katz2023} with a difference of $2$~mag in $G$, we find that the resulting variation in radial velocity corrections can reach up to $0.5~\mathrm{km\,s^{-1}}$ (see Figure~\ref{fig:gaia_magnitudes_comparison}, right panel). To account for the additional uncertainty introduced by using $G$ as a proxy, we compute the correction assuming $G \pm 2$~mag. We add the corresponding difference as an extra systematic uncertainty to the radial velocity errors for those sources without available $G_{\mathrm{RVS}}$ magnitudes (Eq. \ref{eq:corr_e_rv}, where $f_{\text{mag}}$ is our fit, and $f_{\sigma}$ is from eq. 1 of \citealt{Babusiaux2023}).

\begin{align}
    \label{eq:corr_e_rv}
    \text{e\_RV\_corr} = \text{e\_RV} + \sigma, \\
    \sigma = \sqrt{f_{\sigma}^{2} + f_{\text{mag}}^{2}.}
\end{align}

If the used magnitude was the $G_{\mathrm{RVS}}$ magnitude, $f_{\text{mag}} = 0$, and $\sigma$ is reduced to the \cite{Babusiaux2023} correction. The catalog has a dedicated flag, \texttt{G magnitude}, which indicates whether a $G_{\mathrm{RVS}}$ magnitude is available (\texttt{G\_RVS}) or not (\texttt{G}). A small number of stars ($\sim 900$) do not have either $G_{\mathrm{RVS}}$ or $G$ magnitude, so they were removed from the sample.

In addition to the radial velocity adjustments, we also implement the proper motion corrections proposed by \cite{Cantat-Gaudin2021}. 
To assess the impact of these corrections on the derived orbital parameters, we randomly select a sample of approximately $1{,}000{,}000$ \textit{Gaia} stars with $G < 13$, apply the proper motion corrections, and recompute the corresponding orbital parameters. A comparison of orbital parameters obtained with and without proper-motion corrections shows that the differences are consistently lower than $0.61\%$ for the velocities, eccentricity, and actions, demonstrating that the corrections have a negligible effect on the orbital solutions.

\begin{figure*}[htb!]
    \centering
    \includegraphics[width=1\linewidth]{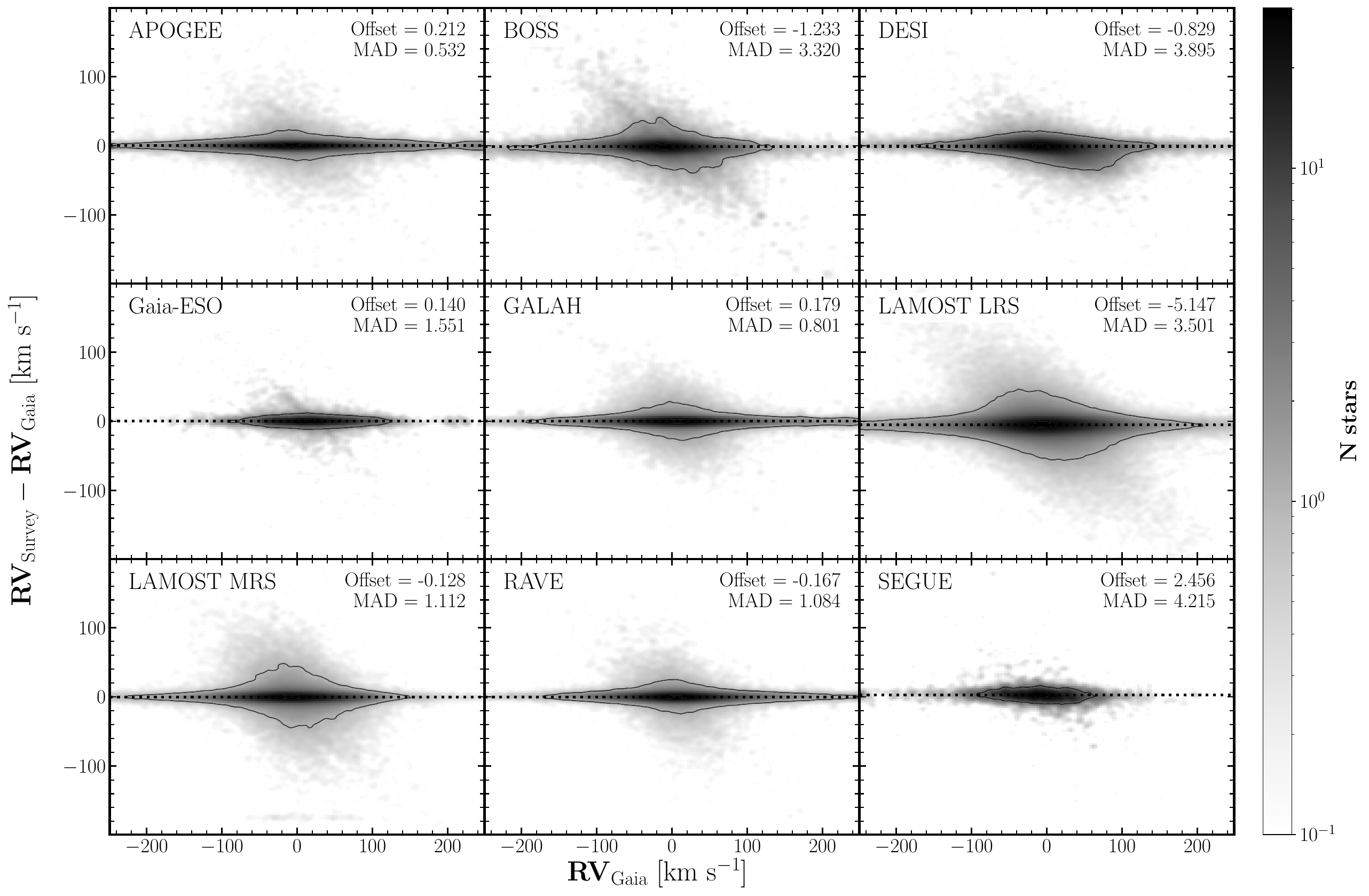}
    \caption{Offset between \textit{Gaia}'s radial velocity and each catalog's radial velocity, in ${\rm km\,s^{-1}}$. The black dashed line represents the offset, computed from the median of all objects, and the surrounding line represents $1\sigma$ of the objects.}
    \label{fig:offsets}
\end{figure*}


\section{Systematic Offsets and Homogenization of Radial Velocities}
\label{sec:rv_norm}

As a previous step, we first calibrate the radial velocities of each catalog to the \textit{Gaia} DR3 reference frame, ensuring consistency across datasets. To derive the radial velocity offsets, we construct a calibration subsample by applying the quality criteria specific to each survey, as detailed below. These quality cuts are used solely to determine the radial velocity offsets and are not applied to the final catalog, which retains all sources.

We use the condition \texttt{RUWE} $<1.4$ to select sources with reliable astrometric measurements \citep{GaiaDR3}. We restrict the effective temperature to the range 3,500--6,500~K to retain unevolved stars and avoid low-quality solutions associated with very cool or very hot stars.

For each survey (APOGEE, BOSS, DESI, \textit{Gaia}-ESO, GALAH, LAMOST LRS, LAMOST MRS, RAVE, and SEGUE), we adopted the radial velocity measurements provided by the original catalogs, using the survey-specific columns ({\tt v\_rad}, {\tt XCSAO\_RV}, {\tt RV\_ADOP}, {\tt VRAD}, {\tt rv\_comp\_1}, {\tt rv}, {\tt rv\_br1}, {\tt HRV}, and {\tt RV\_ADOP}, respectively). For LAMOST MRS, which provides multiple radial velocity estimates, we select the measurement that minimizes both the MAD and the systematic offset relative to \textit{Gaia} radial velocities.

We also notice that radial velocity uncertainties correlate with SNR. To ensure more reliable measurements, we remove the stars within the lowest 5th percentile of SNR in each survey\footnote{SNR thresholds adopted for the 5th-percentile cut: APOGEE: SNR $> 10$; BOSS: SNR $>10$; DESI: SNR $>10$; \textit{Gaia}-ESO: SNR $>5$; GALAH: SNR $>25$; LAMOST LRS: SNR $>15$; LAMOST MRS: SNR $>15$; RAVE: SNR $>20$; SEGUE: SNR $>10$.}, which generally correspond to the largest radial velocity uncertainties. The only exception was APOGEE, for which the 5th-percentile SNR cut would have been overly restrictive, removing stars with small radial velocity uncertainties; therefore, the SNR condition was relaxed. In addition, we apply a uniform cut across all surveys, retaining only stars with radial velocity uncertainties smaller than 10~km~s$^{-1}$, except for RAVE, as discussed below.

Furthermore, we apply specific quality flags to retain only sources with accurate radial velocity measurements. For APOGEE, we keep only stars with \texttt{spectrum\_flags $==$ 0}, removing all objects in APOGEE DR19 with a non-zero bitmask\footnote{\url{https://www.sdss.org/dr19/data_access/bitmasks/}}. For BOSS, we adopt \texttt{ZWARNING $==$ 0} to exclude spectra with known issues. For the DESI Milky Way Survey, since we use the FERRE pipeline, we apply the quality filters recommended for that catalog, which are \texttt{CHISQ\_TOT $< 5$}, \texttt{FEH $> -4.9$}, avoiding the stars that accumulate at the edge of the grids, and \texttt{BESTGRID $!=$ ``s\_rdesi1''}, avoiding the stars that were fitted with the PHOENIX grid. 
For \textit{Gaia}-ESO, we exclude all spectra flagged\footnote{\url{https://www.eso.org/rm/api/v1/public/releaseDescriptions/210}} as \texttt{SRV}, \texttt{SRO}, or \texttt{BIN}, corresponding respectively to suspicious or unreliable radial velocity, suspicious rotational velocity, and identified binaries. For GALAH, we adopt the recommended quality requirement \texttt{flag\_sp $==$ 0}\footnote{\url{https://www.galah-survey.org/dr4/using_the_data/}} and use a slightly lower SNR (\texttt{snr\_px\_ccd3 > 25}), which remains consistent with the SNR criteria discussed above and does not affect the radial velocities. Some GALAH objects lack SNR values; for these cases, we replace the missing {\fontfamily{lmtt}\selectfont NaN} entries with {\fontfamily{lmtt}\selectfont SNR = 0}. For LAMOST LRS and MRS, we retain only stars with reliable radial velocity flags (\texttt{rv\_br\_flag $==$ 0}). For RAVE, we follow the recommendations of \cite{Steinmetz2020} and \cite{Kordopatis2013}, keeping stars that satisfy $|{\rm correctionRV}| < 10~{\rm km\,s^{-1}}$, ${\sigma({\rm RV})} < 8~{\rm km\,s^{-1}}$, and ${\rm correlationCoeff} > 10$. For SEGUE, to avoid spectra with known problems, we adopt \texttt{FLAG $==$ ``nnnnn''}.

 We obtain the radial velocity offsets between each spectroscopic survey and \textit{Gaia}, adopting \textit{Gaia} radial velocities as the reference. The resulting radial velocity offsets are shown in Figure~\ref{fig:offsets}. These offsets, derived from the cleaned sub-samples, are then applied to the entire dataset of each survey, thereby placing all radial velocities on the same scale. This procedure ensures that the radial velocities from the survey catalogs are consistently calibrated and can be used in subsequent analyses. The normalized radial velocities were then used as one of the inputs for the orbit calculation described in the following Section.

Finally, we also examine the radial-velocity differences as a function of stellar and observational parameters to assess the consistency of the calibration across different regions of parameter space. The RV differences remain approximately constant across the main parameter ranges. Furthermore, the derived orbital parameters are not strongly sensitive to RV variations at the level of the largest typical survey dispersion.  The results of this analysis are presented in Appendix~\ref{appendix_param_rv}.


\begin{figure*}[htb!]
    \centering
    \includegraphics[width=0.99\linewidth]{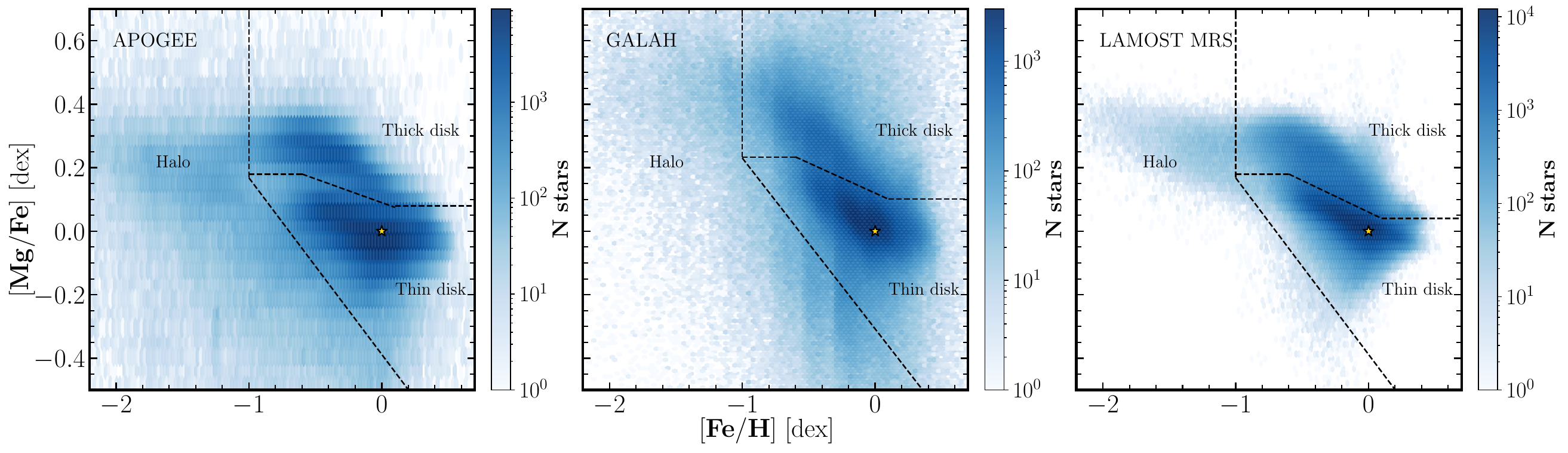}
    \caption{[Fe/H] \textit{vs.} [Mg/Fe] relation for APOGEE, GALAH, and LAMOST MRS. The yellow star indicates the Sun's position, and the dashed lines divide the graphic into the main local Galactic components.}
    \label{fig:tinsley_diagram_surveys}
\end{figure*}

\section{Orbital Parameters}
\label{sec:orb_par}

To compute the orbital parameters of the stars in our sample, we employ an algorithm adapted from \cite{Limberg2021}\footnote{\url{https://github.com/guilimberg/orbit-calc}}, which makes use of the \texttt{AGAMA} code \citep{Vasiliev2019}, 
adopting the axisymmetric Galactic potential model described by \cite{McMillan2017}\footnote{The effect of adding a realistic bar potential will be studied in future work.}. The determination of the orbital parameters is carried out using 100 Monte Carlo integrations (see Appendix \ref{appendix_SEGUE_MC} for further details). In our analysis, the position of the Sun relative to the Galactic Center is assumed to be $(X, Y, Z)_{\odot} = (-8.2, 0.0, 0.0) \, \text{kpc}$, as outlined in \cite{BlandHawthorn2016}. Additionally, the local circular velocity is set to $232.8 \, \text{km s}^{-1}$, consistent with the \cite{McMillan2017} model, while the solar motion with respect to the local standard of rest (LSR) is taken as $(U, V, W)_{\odot} = (11.10, 12.24, 7.25) \, \text{km s}^{-1}$, following the values determined by \cite{schonrich2010}.

The inputs required for orbit computation are the six-dimensional space of positions and velocities: right ascension (RA), declination (DEC), heliocentric distance ($d$), radial velocity (RV), proper motion components ($\mathrm{pmRA}$, $\mathrm{pmDEC}$), and their respective uncertainties. With the stars' 6D phase-space information, we integrate their orbits forward in time for 10 Gyr with \texttt{AGAMA} and calculate their eccentricity ($e$) \footnote{Eccentricity is defined as $e = \frac{R_{\text{apo}} - R_{\text{peri}}}{R_{\text{apo}} + R_{\text{peri}}}$, with $e=0$ being a circular orbit.}, orbital inclination ($i$)\footnote{Orbital inclination is defined as $\cos^{-1}\left(\frac{L_Z}{L}\right)$, where $L_Z$ is the vertical component of the total angular momentum; from this definition, $i < 90^{\circ}$ represents prograde motion.}, and pericentric ($R_{\text{peri}}$) and apocentric ($R_{\text{apo}}$) passage, as well as the Galactocentric radius ($R_{\text{max}}$) and the maximum height above the Galactic plane ($Z_{\text{max}}$). We use the St{\"a}ckel approximation implemented in \texttt{AGAMA} to calculate the radial and vertical actions, $J_r$, $J_z$, respectively. We also provide the Energy ($E$), and the total ($L_{\text{tot}}$) and perpendicular ($L_{\text{perp}}$) angular momentum\footnote{$L_{\text{tot}}$ and $L_{\text{perp}}$ are defined as $L_{\text{tot}} = \sqrt{L_x^2 + L_y^2 + L_z^2}$ and $L_{\text{perp}} = \sqrt{L_x^2 + L_y^2}$, respectively.}. The output also includes the current Galactocentric positions in Cartesian coordinates, velocities in Cartesian ($V_X$, $V_Y$, $V_Z$) and cylindrical ($V_r$, $V_{\phi}$, $V_z$) coordinates, and the Cartesian angular momentum components ($L_X$, $L_Y$, $L_Z$). The 16th and 84th percentiles are provided for all quantities, reflecting the propagated uncertainties from the observational measurements. We refer the reader to Appendix \ref{appendix_columns} for a full description of the input and output parameters.

For a small fraction of the catalog, \texttt{AGAMA} was unable to compute reliable orbital parameters because one or more input quantities contained outlier values (e.g., unusually high proper motions or large radial-velocity uncertainties), preventing the orbit integration from being completed within the adopted 10 Gyr interval. Whenever more than $16\%$ of the integrations failed, the affected orbital parameters were assigned \texttt{NaN}. Nevertheless, since we still have reliable positions and velocities for these objects, we decided to keep them in the catalog. These cases represent a small fraction of our catalog, $\sim 0.03\%$ for SH, and $\sim 0.68\%$ for BJ.


\section{Canonical component classification}
\label{sec:components}

In addition to providing a catalog of kinematic and orbital parameters, we classify stars by their Galactic components using both kinematic and chemical criteria. This classification enables the identification of the canonical stellar populations of the Milky Way, namely the thin disk, thick disk, and halo. A detailed characterization of sub-populations within these components or accreted substructures is beyond the scope of this work.

\subsection{Kinematical classification}

The kinematical classification is based on the probability that a star belongs to a given Galactic component, following the approach presented in \cite{Perottoni2021}. This method builds on the formalism originally introduced by \cite{Bensby2003}, adapted to the cylindrical velocity space $(v_R, v_{\phi}, v_z)$. In this framework, the probability distribution function follows the velocity ellipsoid prescription given in Eq.~B1 of \cite{Perottoni2021}.  To account for the radial variation of Galactic rotation, we compute the azimuthal velocity relative to the circular velocity at the star's Galactocentric radius, adopting the rotation curve of \citet{Eilers2019}.

The parameters defining the velocity ellipsoids for the thin disk, thick disk, and halo are adopted from \cite{Kordopatis2013} and \cite{Amarante2021} and are summarized in Table \ref{tab:bensby2003}, where
$V_{\text{asym}}$ is the asymmetric drift and $\sigma_{V_i}$ refers to the characteristic velocity dispersions in the Galactic component $i$. F is the fraction of stars observed for a given population in the Solar neighborhood. We compute membership likelihoods for each star and use ratios of probabilities of belonging to different Galactic components to assign stars to the corresponding populations.

As a result, we include three columns in the catalog that quantify the kinematic membership probabilities for the thin disk ($P\_D\_K$), thick disk ($P\_TD\_K$), and halo ($P\_T\_H$). These probabilities provide a quantitative measure of the likelihood that each star belongs to a given Galactic component based on its kinematic properties.

We caution, however, that the adopted kinematic classification is based on velocity ellipsoids calibrated in the Solar neighborhood and should therefore be interpreted with care for stars at large heliocentric distances. Although our implementation accounts for the radial variation of the Galactic circular velocity, it does not account for possible changes in the velocity ellipsoids with Galactic position (e.g. \citealt{Sharma2020, Sun2023, Sun2025}). We therefore provide kinematic membership probabilities for all stars in the catalog, but recommend that they be interpreted with caution outside the regime in which the velocity ellipsoids have been calibrated.

\begin{deluxetable}{llllll}
\tablecaption{Parameters for the ellipsoid velocity for each Galactic component at the Solar Neighborhood. \label{tab:bensby2003}}
\tablehead{
\colhead{ } & \colhead{F} & \colhead{$\sigma_{V_r}$} & \colhead{$\sigma_{V_\phi}$} & \colhead{$\sigma_{V_Z}$} & \colhead{$V_{\text{asym}}$} \\
\colhead{} & \colhead{} & \colhead{} & \colhead{[\%]} & \colhead{} & \colhead{[km s$^{-1}$]}
}
\startdata
Thin disk (D)   & 0.9301 & 30  & 20  & 18  & $-$28  \\
Thick disk (TD) & 0.0652 & 61  & 45  & 44  & $-$63  \\
Halo (H)        & 0.0047 & 160 & 119 & 110 & $-$228 \\ \hline
\enddata
\end{deluxetable}

\subsection{Chemical classification}
\label{subsec:chemical_class}

The chemical classification is based on the [Mg/Fe] vs. [Fe/H] diagram, introduced by \cite{Wallerstein1962} and \cite{Tinsley1980}, which links elemental abundance patterns to the chemical evolution of the Milky Way and enables the separation of its main components into the thin disk, thick disk, and halo. We apply this classification to stars observed by APOGEE, GALAH, and LAMOST, the surveys that provide sufficiently precise and homogeneous abundance measurements, allowing us to distinguish the Galactic populations in the [Mg/Fe]--[Fe/H] plane.

The classification is based on simple, survey-specific selection functions defined by piecewise linear relations in the [Mg/Fe]--[Fe/H] plane (see Figure~\ref{fig:tinsley_diagram_surveys}). 
These relations delineate the low-$\alpha$, high-$\alpha$, and low-metallicity (stellar halo; \citealt{Fuhrmann1998}) populations and are chosen to follow the observed bimodality in each survey. Stars lying above the defined boundary are classified as high-$\alpha$, while those below are classified as low-$\alpha$. 

 To avoid unnecessarily restricting the combined sample, we define similar, although not identical, selection boundaries in the [Mg/Fe]--[Fe/H] plane for each survey (see Appendix~\ref{appendix_chemical_selection} for details). We use this single abundance space for all three surveys, as they do not provide the same set of elemental abundances, and not all stars have measurements for every element. Although any selection in chemical space inevitably results in some level of contamination between populations, these selection criteria provide a practical way to identify the main Galactic components \citep{Alinder2025}.

\begin{deluxetable*}{lcccccccc}
\tablecaption{Number of objects for each survey. \label{tab:results}}
\tablehead{
\colhead{} &
\multicolumn{4}{c}{BJ distances} &
\multicolumn{4}{c}{SH distances}\\
\noalign{\vskip -5pt}
\colhead{} &
\multicolumn{4}{c}{\rule{7.5cm}{0.1pt}} &
\multicolumn{4}{c}{\rule{7.5cm}{0.1pt}}\\
\noalign{\vskip -5pt}
\colhead{Survey} &
\colhead{Full sample} &
\colhead{Unique} &
\colhead{Clean Sample} &
\colhead{Golden Sample} &
\colhead{Full sample} &
\colhead{Unique} &
\colhead{Clean Sample} &
\colhead{Golden Sample}
}
\startdata
APOGEE DR19             &  1,074,161    &  955,232     & 498,797      &  461,776           &  846,575      & 746,459      & 409,943        &  377,172     \\
BOSS DR19               &  815,485      &  274,277     & 350,912      &  112,914           &  545,180      & 185,150      & 297,121        &  93,643      \\
DESI DR1                &  4,745,455    &  3,865,849   & 2,351,440    &  1,945,453         & 1,846,897     & 1,287,670    & 1,223,080      & 943,647      \\
\textit{Gaia} RVS DR3   &  33,581,727   &  30,684,239  & 33,581,727   &  31,763,203        & 26,660,281    & 24,111,678   & 26,660,281     & 25,035,402   \\
\textit{Gaia}-ESO DR5   &  109,333      &  104,998     & 77,719       &  75,601            & 53,993        & 52,368       & 41,787         & 41,000       \\
GALAH DR4               &  1,062,120    &  838,425     & 569,012      &  480,996           & 960,177       & 766,063      & 528,859        & 447,646      \\
LAMOST LRS DR11         &  7,689,717    &  2,722,296   & 5,454,828    &  1,847,182         & 6,603,183     & 2,273,287    & 4,744,436      & 1,568,565    \\
LAMOST MRS DR11         &  2,533,830    &  1,026,725   & 1,591,501    &  681,163           & 2,279,215     & 921,330      & 1,443,394      & 616,351      \\
RAVE DR6                &  444,006      &  376,358     & 301,455      &  269,027           & 400,646       & 340,068      & 275,086        & 245,751      \\
SEGUE DR12              &  367,399      &  252,467     & 197,602      &  139,586           & 159,422       & 93,060       & 107,147        & 67,931       \\  
Total                   &  52,423,233   &  41,100,866  & 44,974,993   &  37,776,901        &  40,355,569   & 30,777,133   & 35,731,134     & 29,437,108   \\ \hline
\enddata

\end{deluxetable*}


 \begin{figure*}[htb!]
    \centering
    \includegraphics[width=\linewidth]{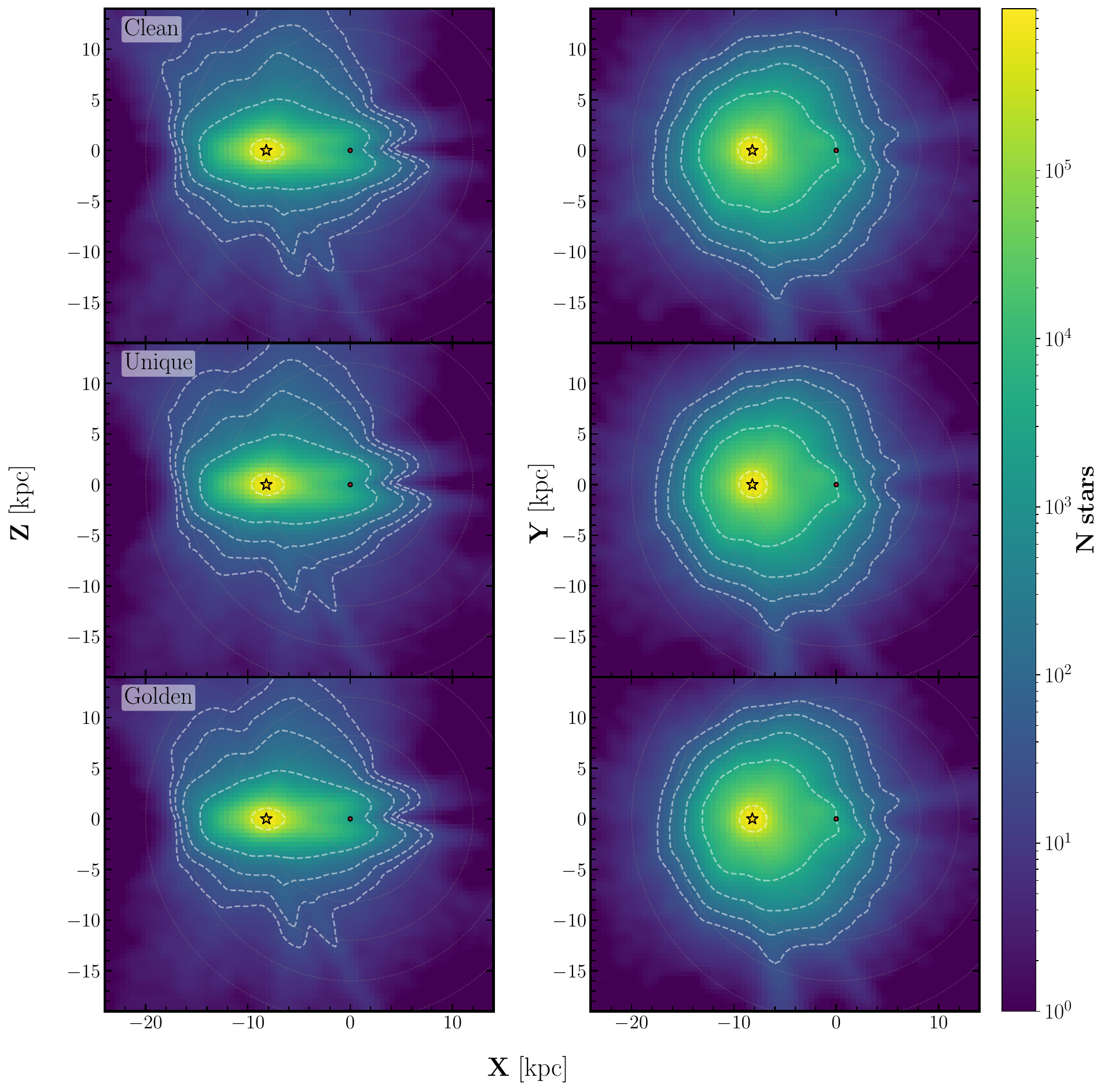}
    \caption{Spatial distribution for the final catalog in Galactocentric coordinates: the \texttt{Clean} catalog (upper panel), \texttt{Unique} (central panel), and \texttt{Golden} (lower panel). The yellow star-shaped dot represents the Sun's position relative to the Galactic center, marked with the maroon dot.}
    \label{fig:xyz_master}
\end{figure*}

\begin{figure*}[htb!]
    \centering
    \includegraphics[width=\linewidth]{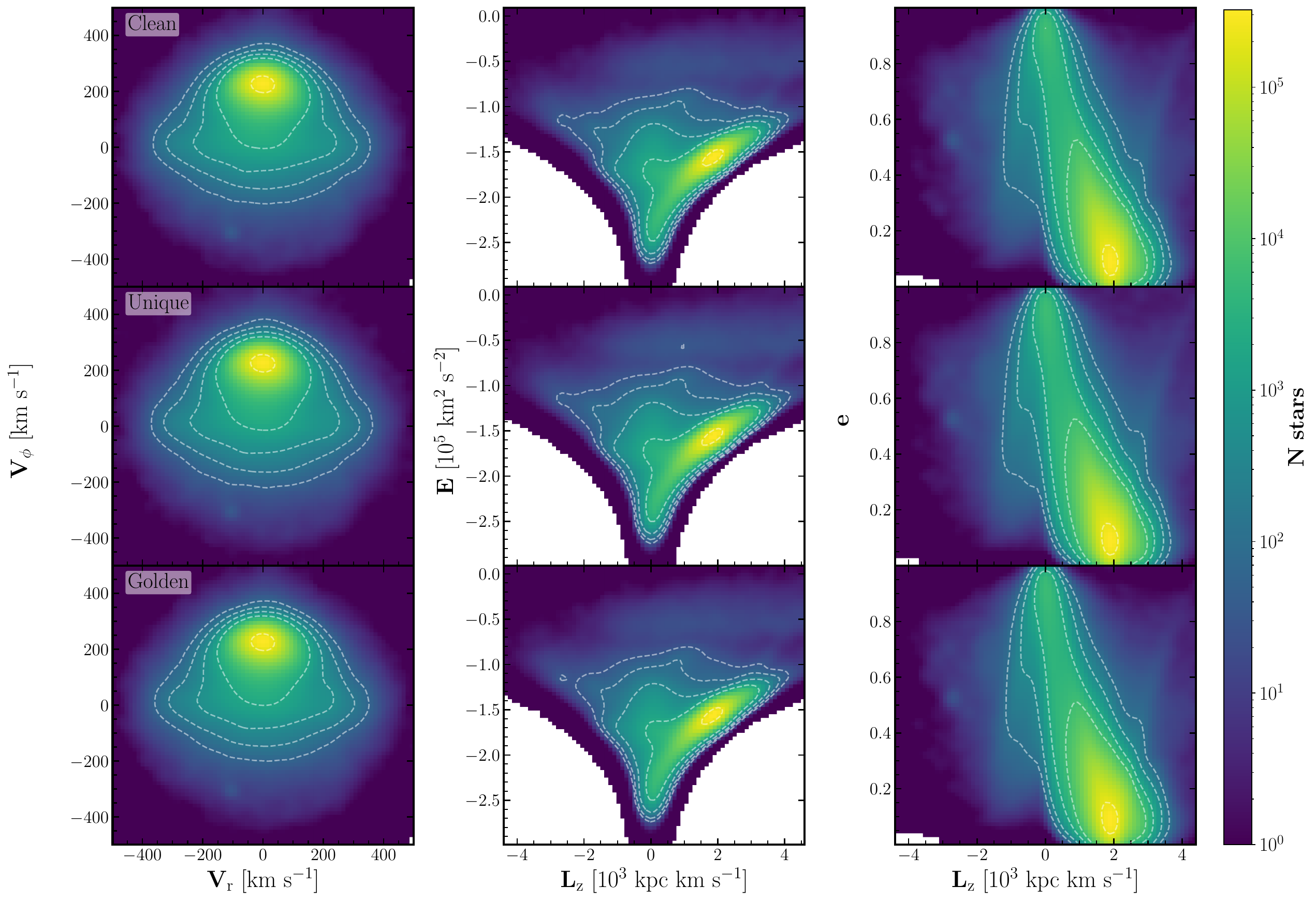}
    \caption{Orbital parameters for the final catalog: $V_r$ \textit{vs.} $V_{\phi}$ (left panel), $L_Z$ \textit{vs.} $E$ (central panel), and $LZ$ \textit{vs.} $e$ (right panel); the superior panels refer to the \texttt{Clean} catalog, the central panels refer to the \texttt{Unique} sample, and the lower panels refer to the \texttt{Golden} sample.}
    \label{fig:kinematics_master}
\end{figure*}

\begin{figure}[htb!]
    \centering
    \includegraphics[width=0.95\linewidth]{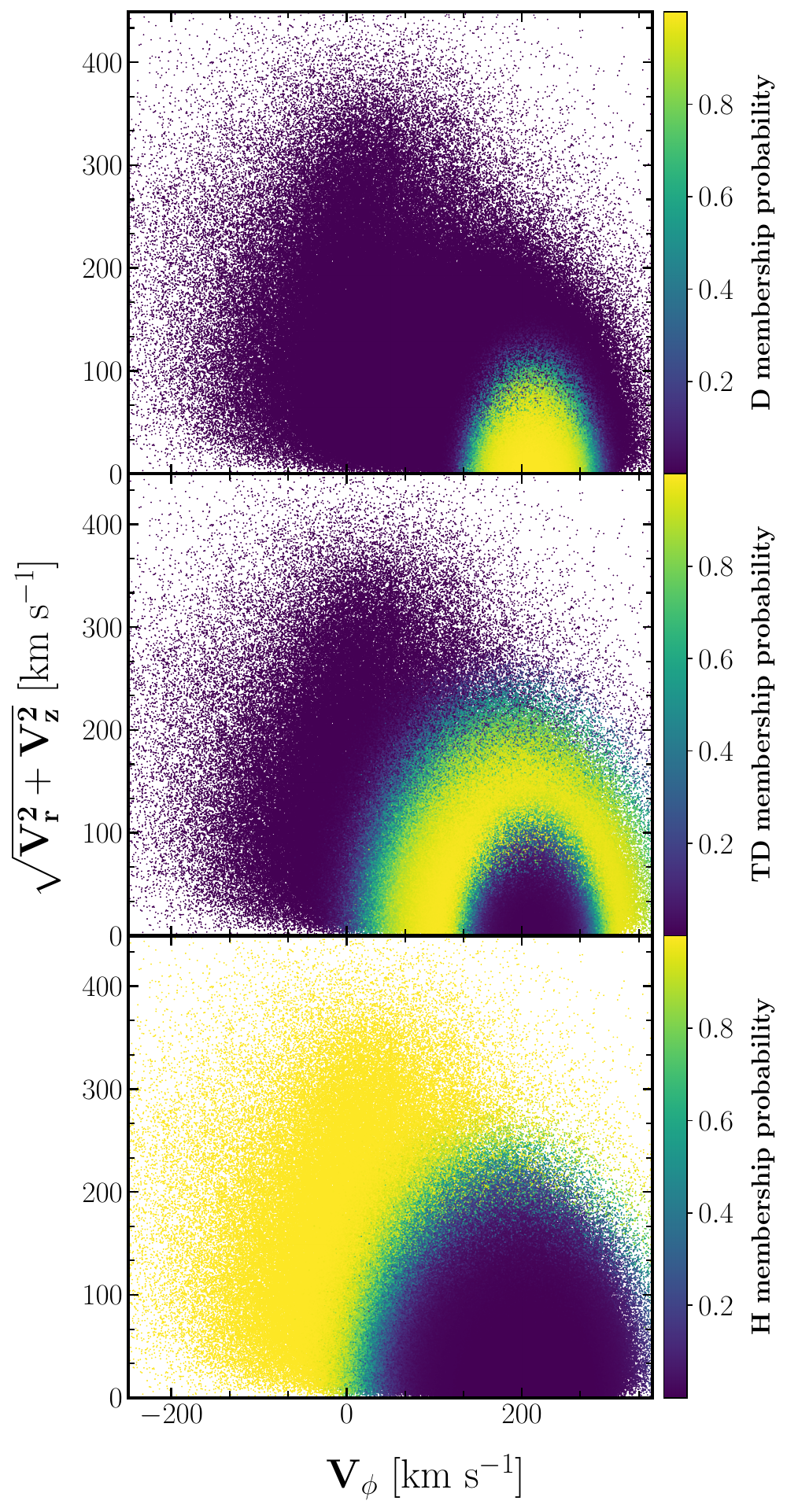}
    \caption{Toomre diagram for SH \texttt{Golden} sample, color-coded according to their probability of belonging to the thin disk (upper panel), thick disk (middle panel), and halo (lower panel). The color bar indicates the membership probability for each Galactic component.}
    \label{fig:toomre_catalog_prob}
\end{figure}

\begin{figure*}[htb!]
    \centering
    \includegraphics[width=\linewidth]{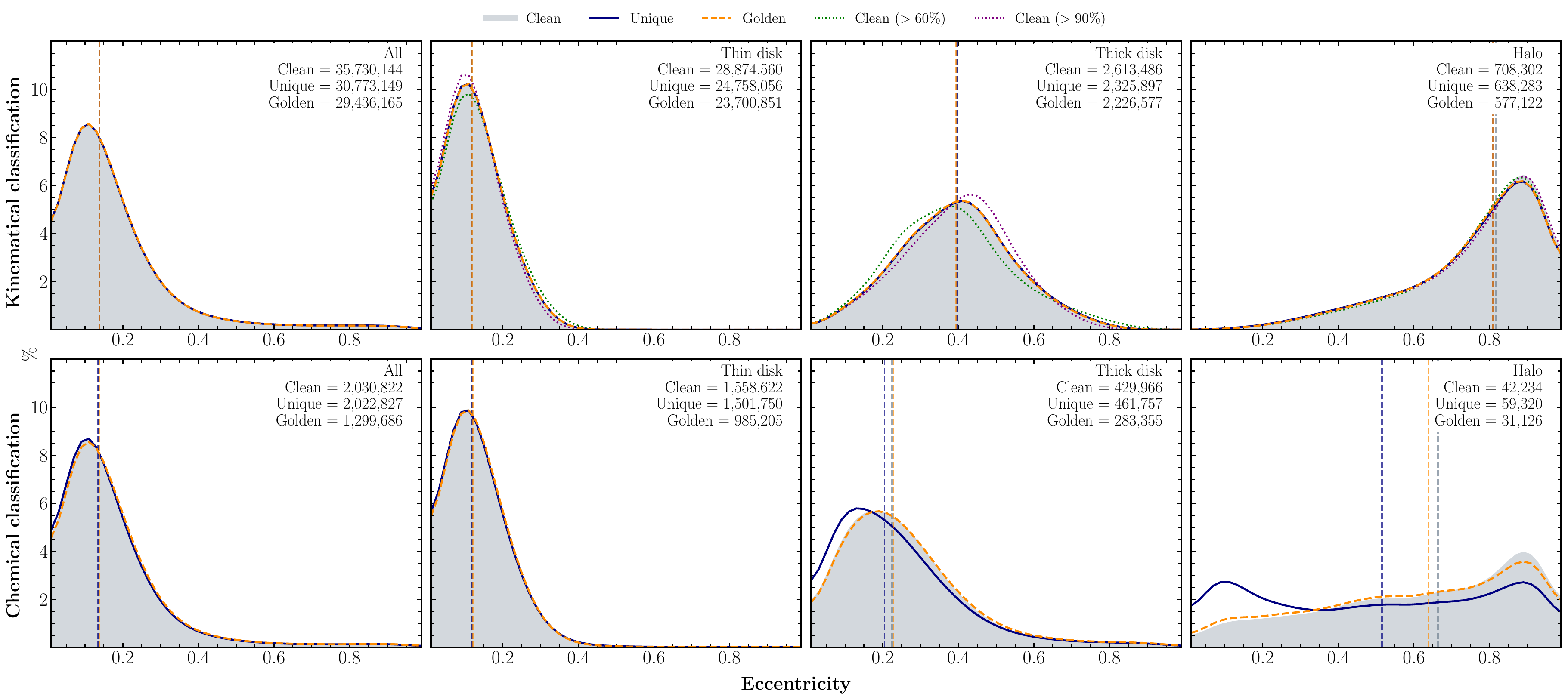}
    \caption{Upper panels: Eccentricity distributions for stars classified according to their kinematic membership probabilities in different Galactic regions. The \texttt{Clean} sample, shown by the filled gray distribution, includes stars with kinematic membership probabilities up to 80\%. The \texttt{Unique} and \texttt{Golden} samples are shown by the blue solid and orange dashed lines, respectively. Lower panels: Eccentricity distributions for stars classified according to their chemical membership in different Galactic regions, using the same samples. In both panels, the green and purple dotted lines show the \texttt{Clean} sample distributions for probability thresholds of 60\% and 90\%, respectively. Vertical dotted lines indicate the median eccentricity of each distribution. The number of stars in each sample is indicated in the upper-right corner of each panel.}
    \label{fig:hist_master}
\end{figure*}

\section{Catalog Contents}\label{sec:catalog}

The \texttt{Full} catalog is constructed to include all objects, regardless of data quality or individual measurement precision. We opt to retain all stars (i.e., all objects from each survey without the quality cuts described in Section \ref{sec:obs}), as they may still be useful for specific scientific applications. For each object, we preserve the survey name and original survey identifier, ensuring easy identification and cross-matching, and allowing users to trace back to the original surveys for additional information not included in our catalog. The identifier columns adopted from original surveys were: DESI: {\tt TARGETID}; \textit{Gaia}: {\tt source\_id}; \textit{Gaia}-ESO: {\tt OBJECT}; GALAH: {\tt sobject\_id}; LAMOST LRS and MRS: {\tt obsid}; RAVE: {\tt obsID}; SEGUE: {\tt SPECOBJID}. Since SDSS-V does not provide a truly unique identifier for individual observations i.e., when a star has multiple observations, there is no direct way to distinguish them, we constructed a survey identifier based on the SNR. Then, the APOGEE and BOSS survey identifiers in our catalog are {\tt sdss\_id.SNR}. If the user needs to recover the original SDSS identifier, they can extract this information before the first dot in the \texttt{Survey\_ID} column.

The catalog contains both the input and output quantities described in Section~\ref{sec:orb_par}, together with the probabilities of membership in each Galactic component (see Section \ref{sec:components}). We also retain the original signal-to-noise ratio from the spectroscopic surveys. To facilitate the selection of higher-quality subsets, we further define three flags: \texttt{Clean}, \texttt{Unique}, and \texttt{Golden}.

The \texttt{Clean} flag identifies stars with highly reliable radial velocities, based on the same criteria adopted for the radial velocity normalization (see Section~\ref{sec:rv_norm}). For the three flags, entries satisfying the corresponding criteria are marked as \texttt{TRUE}, while the remaining entries are marked as \texttt{FALSE}.

The \texttt{Unique} flag identifies the preferred entry for stars with multiple measurements. This selection follows a hierarchy that accounts for both data quality and quantity: APOGEE, GALAH, LAMOST MRS,  \textit{Gaia}-ESO, RAVE, \textit{Gaia}, LAMOST LRS, DESI, BOSS, and SEGUE. When multiple observations are present in the same survey, the entry with the highest signal-to-noise ratio is selected. The adopted hierarchy directly affects the \texttt{Unique} flag. 

The \texttt{Golden} flag identifies a sample containing only observations that satisfy the \texttt{Clean} criteria. Then, among the observations satisfying these criteria, a single entry per object is selected according to the survey hierarchy adopted for the \texttt{Unique} flag.

Although our analysis focuses on stars bound to the Milky Way, external systems are naturally included because they are present in the original surveys. For this reason, we include a flag to identify stars associated with the Magellanic Clouds, globular clusters, Sagittarius, and dwarf galaxies, identified by cross-matching our catalog with the compilations presented by \cite{Nidever2020}, \cite{Vasiliev2021}, \cite{Ramos2022}, and \cite{Pace2022}, respectively. The \texttt{Population} flag labels \texttt{DG} for dwarf galaxy members, \texttt{LMC} and \texttt{SMC} for Magellanic Cloud members, \texttt{Sgr} for Sagittarius members, and \texttt{GC} for globular cluster stars. Nevertheless, they constitute only a small fraction of the catalog, comprising $98{,}166$ objects for Bailer-Jones and $55{,}902$ objects for Star-Horse distances ($\sim0.19\%$ and $\sim0.13\%$  of the \texttt{Full} sample, respectively). Objects associated with the Milky Way potential are flagged as \texttt{MW}.

Overall, the catalog is dominated by thin-disk stars, resulting in their larger contribution to the global distributions (see Section \ref{sec:results}). Halo stars are the most strongly affected by the quality selections: only $49\%$ are retained in the \texttt{Clean} sample, compared to $\sim73\%$ of thin- and thick-disk stars. This reflects the combined impact of larger distance uncertainties and lower signal-to-noise ratios typical of halo targets.

We refer the reader to Appendix \ref{appendix_columns} for a description of the columns of our catalog. The final catalog contains 52,423,233 stellar sources for Bailer-Jones distances, and 40,355,569 for \texttt{StarHorse} distances. Among them,
the total number of sources flagged as ``\texttt{TRUE}'' in the \texttt{Unique} category is  41,100,866 and 30,777,133,  
 44,974,933 and 35,731,134 objects are flagged as ``\texttt{TRUE}'' in the \texttt{Clean} flag, and
 37,776,901 and 29,437,108 objects are flagged as ``\texttt{TRUE}'' in the \texttt{Golden} flag, for Bailer-Jones and \texttt{StarHorse}, respectively. The data are summarized in Table \ref{tab:results}.


\section{Results}
\label{sec:results}

The spatial distributions of the subsamples flagged as \texttt{Clean}, \texttt{Unique} and \texttt{Golden}, derived using the SH distances, are shown in Figure \ref{fig:xyz_master}. A feature of these maps is the presence of overdensities towards specific directions, which arise primarily from the surveys' observational strategies, interstellar extinction, and distance uncertainties. For example, the Large and Small Magellanic Clouds (LMC and SMC) appear elongated along their respective lines of sight, within the ranges $-6 < X < 2$~kpc, $-30 < Y < -4$~kpc, and $-15 < Z < 0$~kpc, due to confusion with Milky Way foreground stars (e.g., \citealt{Anders2019}).

\subsection{Kinematic and Dynamical diagrams}

Figure~\ref{fig:kinematics_master} presents examples of the kinematic and dynamical space covered in our catalog. In addition to Figure \ref{fig:toomre_catalog_prob}, which illustrates the distribution of \texttt{Golden} sample stars (see Section \ref{subsec:chemical_class}) in the Toomre diagram, color-coded according to their kinematic membership probabilities, the velocity space defined by the radial velocity in the Galactic plane ($v_R$) and the azimuthal velocity ($v_\phi$) (left column of Figure \ref{fig:kinematics_master}), provides a complementary view of the Galactic components. Across all panels, a prominent concentration of stars is visible around $(v_\phi, v_R) \simeq (240,0)~\mathrm{km~s^{-1}}$, corresponding mostly to the thin disk. Halo stars, in contrast, populate a broad region in $v_R$ and are characterized by low net cylindrical rotation ($v_\phi \sim 0$).

The middle column of Figure \ref{fig:kinematics_master} shows the distribution of stars in the Lindblad diagram: the plane defined by the vertical component of the angular momentum ($L_Z$) and the orbital energy ($E$). Disk stars dominate the region of high prograde angular momentum, whereas halo stars span a wider range of energies and angular momenta, including both prograde and retrograde orbits. More tightly bound stars at lower energies ($E \lesssim -2 \ [\times 10^{5}~\mathrm{km^{2}~s^{-2}}]$) are typically associated with the inner Galaxy. A concentration around $L_Z \sim \ 0 \ [\times 10^{3}~\mathrm{kpc~km~s^{-1}}]$ and $E \sim -1.5 \ [\times 10^{5}~\mathrm{km^{2}~s^{-2}}]$ is consistent with debris from the Gaia-Enceladus/Sausage merger event (GSE; \citealt{Helmi2018,Belokurov2018,Haywood2018,Mackereth2019}), while stars on more retrograde orbits are generally linked to other accreted populations (e.g. \citealt{Naidu2020, Limberg2021}).

The third column of Figure \ref{fig:kinematics_master} displays the $L_Z$–eccentricity plane. Disk stars are characterized by high angular momentum and low eccentricities, consistent with nearly circular orbits. Halo stars occupy a broader region in this space, spanning a wide range of eccentricities and angular momenta, including both prograde and retrograde orbits. The GSE population is particularly prominent as a concentration of stars with high eccentricities ($e \gtrsim 0.7$) and $L_Z \sim 0\ [\times 10^{3}~\mathrm{kpc~km~s^{-1}}]$ (e.g. \citealt{Buder2022, Limberg2022, Lane2023, Carrillo2024}). Several overdensities of retrograde stars with negative $L_Z$ and varying eccentricities indicate the contribution of (small) accretion events to the present-day stellar halo \citep{Review_galacticarchaeology2020, Review_galacticarchaeology2024}.

When individual surveys are examined separately (Appendix \ref{appendix_surveys}), the different sky coverages are clearly visible. GALAH, \textit{Gaia}-ESO, LAMOST MRS, and RAVE are dominated by stars with thin-disk-like dynamics and relatively low dispersion in orbital parameters. In contrast, DESI, \textit{Gaia}, LAMOST LRS, and SEGUE exhibit broader distributions in energy and angular momentum, reflecting a stronger contribution from halo populations. APOGEE shows pronounced overdensities in kinematic space that correspond to explicitly targeted stellar populations, such as globular clusters and nearby dwarf galaxies.

\subsection{Eccentricity distributions}

Figure \ref{fig:hist_master} shows the eccentricity distribution for the \texttt{Clean}, \texttt{Unique}, and \texttt{Golden} samples selected through the kinematical and chemical classification illustrated in Figure \ref{fig:toomre_catalog_prob}. For the kinematical analysis, we define a star as confidently belonging to a certain Galactic component if its membership probability for that component exceeds 80\%. In the top panels, we show the eccentricity distribution for different samples selected through the kinematic selection.  

For chemical classification, we adopt the surveys' flags for [Fe/H] and [Mg/Fe] abundances, as described in Appendix \ref{appendix_chemical_selection}. The eccentricity distributions obtained from the chemical classification for the subset of stars in APOGEE, GALAH, and LAMOST MRS (Figure \ref{fig:tinsley_diagram_surveys}) show similar behavior for the thin disk, but differ significantly for the thick disk and halo components. This likely reflects the known systematic differences between the chemical and kinematic definitions of the thin and thick disks (e.g. \citealt{Bensby2003, Alinder2025}), and the fact that chemical selection does not explicitly separate stellar populations that have undergone substantial dynamical heating, such as the \textit{Splash} population \citep{Bonaca2017, DiMatteo2019, Belokurov2020} or dynamically heated old disk stars. As a result, the chemically selected thick disk displays a pronounced high-eccentricity tail, while the chemically selected halo contains a non-negligible fraction of stars on low-eccentricity orbits. Nevertheless, the fraction of chemically selected halo stars with low eccentricities diminishes significantly when restricting ourselves to the {\tt Golden} sample.

We caution that the kinematic criteria used in our classification naturally imprint a characteristic eccentricity distribution. Selection criteria based on the cylindrical velocity operate as selectors of the total peculiar motion with respect to the Local Standard of Rest (LSR), whereas in the low-to-moderate eccentricity regime, epicyclic theory predicts that curves of constant eccentricity in the $V_X$--$V_Y$ plane are approximately ellipsoidal and defined by a weighted quadratic combination of the velocity components (see, e.g., Fig. 11 of \citealt{Chiba1998}). 
By defining the populations through a restricted range of peculiar motions, our kinematic classification preferentially selects stars that occupy a relatively narrow interval of orbital eccentricities. Due to the quadratic relationship between velocities and eccentricity, this kinematic cut acts as a filter that isolates stars with similar orbital properties. This relationship between orbital eccentricity and peculiar motion in the Toomre Diagram is also clearly evident in Fig. 11 of \cite{Magrini2022}, as well as in our Figure \ref{fig:toomre_catalog_prob}.

Consequently, the eccentricity distribution of a sample classified through the Toomre Diagram is at least partly shaped by the adopted velocity-space selection criteria. This effect helps explain the differences between the eccentricity distributions obtained from the kinematic and chemical classifications. While chemical classification allows stars with different orbital properties to be associated with the same enrichment history, kinematic classification selects stars according to their location in velocity space and therefore tends to recover a narrower range of eccentricities.

\subsection{Mapping kinematic substructures in the disk}

\begin{figure*}[htb!]
    \centering
    \includegraphics[width=0.49\linewidth]{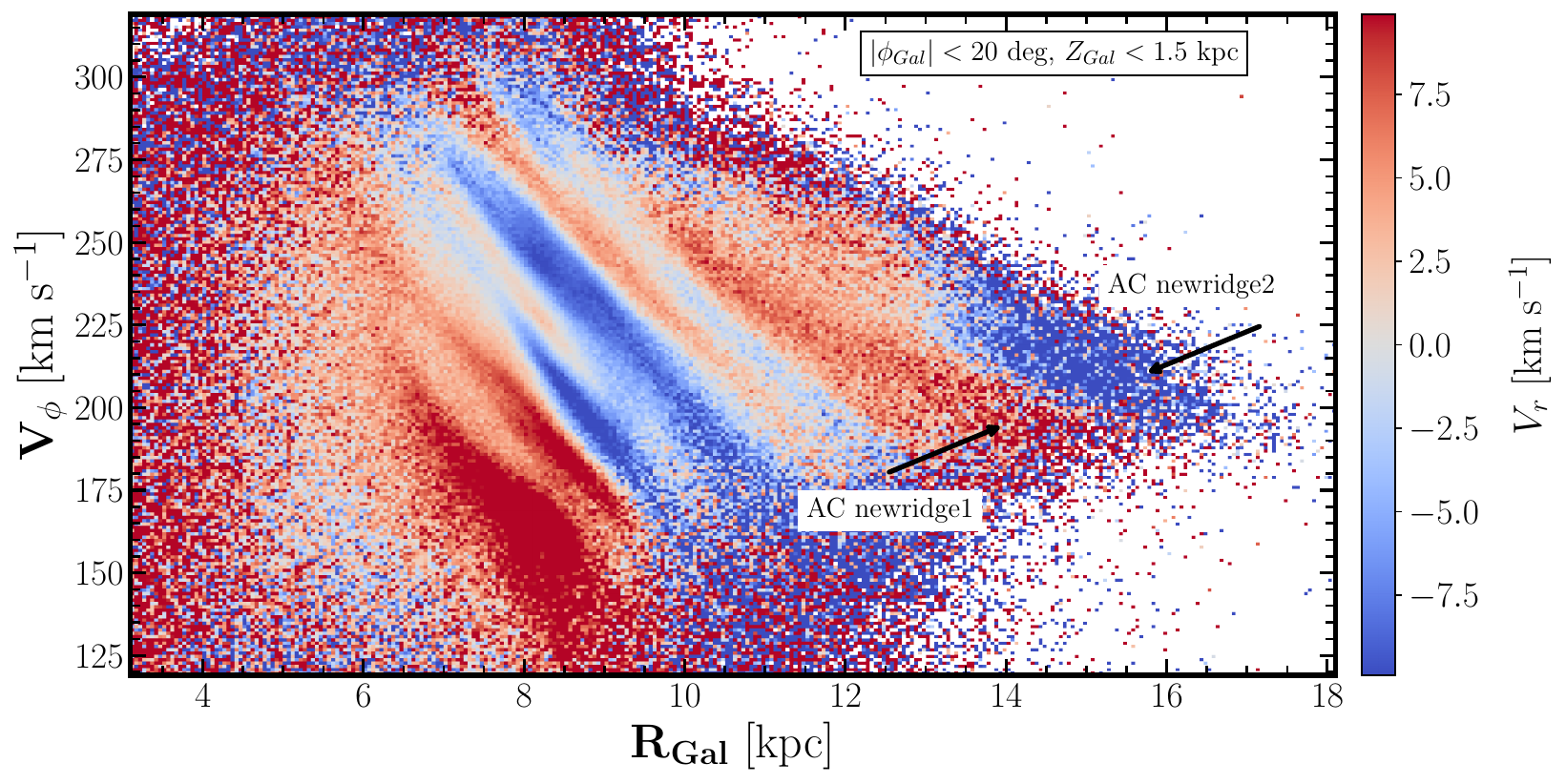}
    \includegraphics[width=0.49\linewidth]{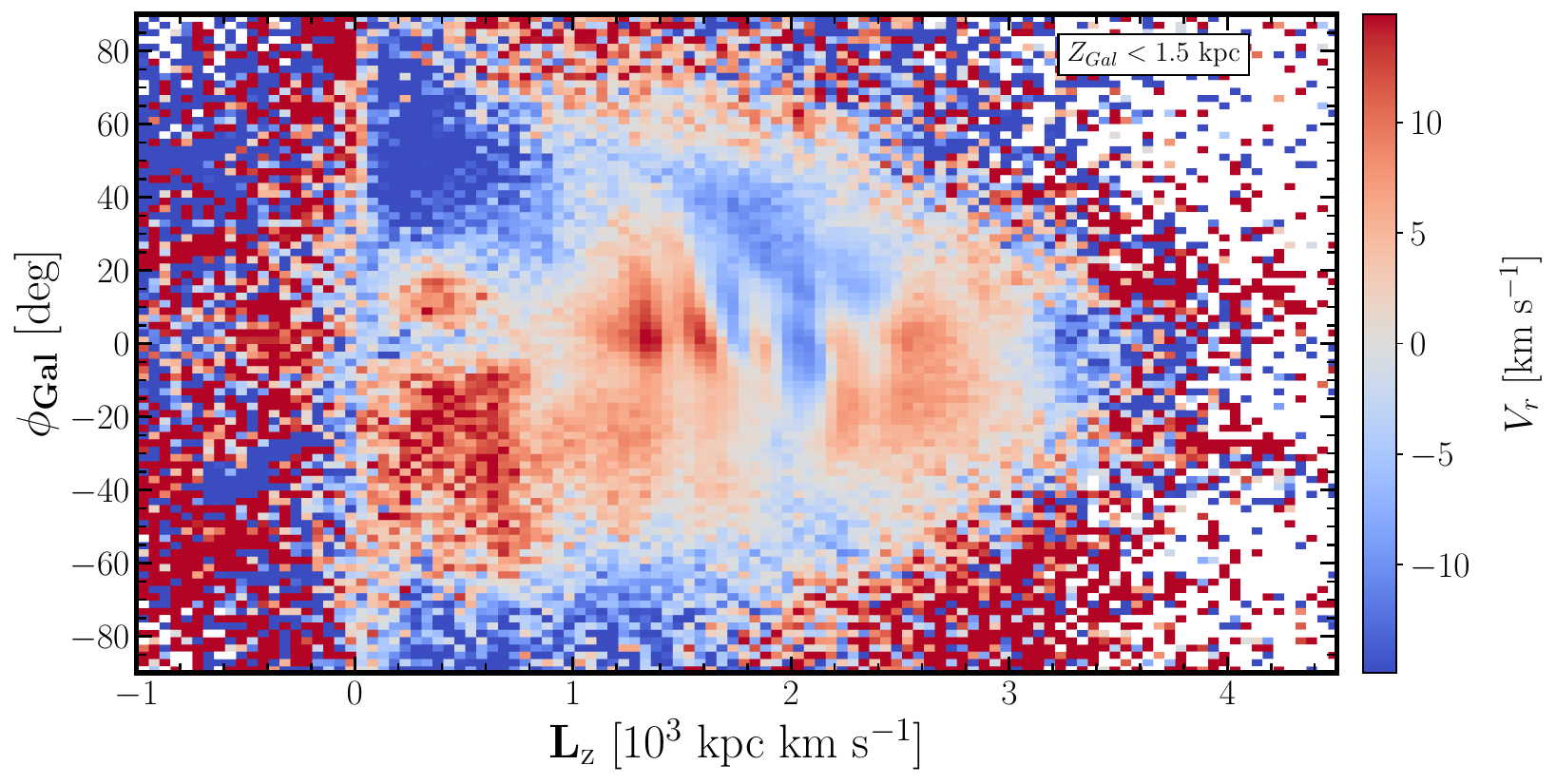}
    \caption{Kinematic ridges and waves in the Milky Way disk ($|Z_{\rm Gal}|<1.5$ kpc) visible in our {\tt Golden} sample. Left panel: Ridges in the $V_{\rm \phi}$ vs. Galactocentric distance plane (|$\phi|<20$ deg; each pixel is color-coded by the mean Galactocentric radial velocity $V_R$). Right panel: The $V_R$ wave in Galactic azimuth $\phi_{\rm Gal}$ vs. angular momentum $L_Z$.}
    \label{fig:antoja22}
\end{figure*}

Figure \ref{fig:antoja22} exemplifies the use of our catalog for the confirmation of, and search for, substructure in the kinematics of the Milky Way disk, in particular the so-called kinematic ridges (\citealt{Antoja2018, Kawata2018, Ramos2018}). The extended spatial coverage of our catalog with respect to the largest single catalog, {\it Gaia} DR3, becomes evident when comparing these plots with their respective counterparts in \citet[][figs. 24c and 25]{Antoja2022}, which were based on 25 million stars with DR3 RVS radial velocities and {\tt StarHorse} distances from \citet{Anders2022}. Our new plots corroborate the substructures detected in those plots, while extending the maps out to $\sim$15 kpc from the Galactic center, reaching regions that were previously dominated by noise.

The left panel of Figure \ref{fig:antoja22} shows the median Galactocentric $V_R$ as a function of Galactocentric distance and rotational velocity, restricted to a wedge of $|\phi_{\rm Gal}|<20$ deg in the Galactic plane. Many of the diagonal ridges in this plot correspond to long-known moving groups of the solar vicinity (e.g. \citealt{Asiain1999, Antoja2008}) whose position in velocity space varies with Galactocentric radius (e.g. \citealt{Ramos2018, Barros2020, Khoperskov2022, Bernet2022, Bernet2024}). The inclination of the ridges depends on the mechanism responsible for their appearance --- some of the structures have been associated with resonances of the bar, with spiral arms, and with the response of the Galactic disk to the pericenter passage of the Sagittarius dSph galaxy (see \citealt{Hunt2025} for a review). 

In the outermost region of the disk, we identify two structures consistent in both position and radial velocity with the Anticenter Ridge 1 and 2 identified by \citet{Antoja2018}. The Anticenter Ridge 2 may represent the low-latitude continuation of the Monoceros Ring \citep{Newberg2002, Xu2015, Borbolato2024}, which extends to higher Galactic latitudes. The negative radial velocity measured for this structure by \citet{Lambert2026} using DESI DR2 led those authors to interpret it as a tidally induced arm, following the model proposed by \citet{Antoja2022}. 

The right panel of Figure \ref{fig:antoja22} maps the $V_R$ waves over a range of angular momenta and Galactic azimuth. This plot was first introduced in \citet{Friske2019} and is a different representation of the ridges. Each of the structures visible in the top panel of Figure \ref{fig:antoja22} corresponds to a same-color overdensity in the lower panel, in the slice $-20\deg < \phi_{\rm Gal} < + 20\deg$. Figure \ref{fig:antoja22} illustrates the potential of our catalog for follow-up kinematic studies, as in this case the larger area covered may allow us to improve our understanding of the dynamical processes causing these ridges (e.g. \citealt{Cao2024, Bernet2025, Hamilton2026, Lambert2026}).


\section{Conclusions}
\label{sec:conclusion}

We present a homogeneous stellar catalog that combines multiple large spectroscopic surveys with \textit{Gaia} DR3 astrometry to derive consistent kinematic, dynamical, and orbital parameters for a broad range of Galactic populations. By leveraging the complementary strengths of different surveys, our catalog achieves improved spatial and kinematic coverage, enabling a more complete description of the Milky Way’s thin disk, thick disk, and halo.

To accommodate different scientific use cases, we provide three complementary samples: the \texttt{Full} catalog, which maximizes completeness; the \texttt{Clean} sample, which prioritizes high-quality astrometry and radial velocities; the \texttt{Unique} sample, which mitigates duplication by selecting a single preferred measurement per star; and the \texttt{Golden} sample, which applies the same quality criteria from the \texttt{Clean} sample while retaining only one observation per star. While these quality-driven selections naturally reduce the representation of halo stars, they ensure robust measurements for more detailed analyses. 

In addition, the catalog includes both kinematic and chemical classifications, providing membership probabilities for the main Galactic components based on stellar dynamics and abundance patterns. We also include dedicated flags to identify stars associated with globular clusters, dwarf galaxies, and the Magellanic Clouds, enabling separation of Milky Way field stars from external systems.

The catalog is made publicly available in Zenodo: \url{10.5281/zenodo.23019541}, with documentation of its contents. We expect it to serve as a flexible reference dataset for future studies of the Milky Way, particularly in the context of upcoming spectroscopic surveys and improved astrometric data releases. Identifying new stellar populations and improving the completeness of known populations helps better understand their nature, contributing to Galactic archaeology. 


\begin{acknowledgments}
N.I.F.S. thanks all those involved with the multi-institutional \textit{Milky Way BR} Group for the weekly discussions and the {\tt StarHorse} team for providing their newly computed distances. H.D.P. thanks FAPESP (proc. 2024/17850-1) and acknowledges the Conselho Nacional de Desenvolvimento Científico e Tecnológico (CNPq) for the support provided through the Research Productivity Fellowship (Grant No. 309431/2026-3). D.S. acknowledges support from the Foundation for Research and Technological Innovation Support of the State of Sergipe (FAPITEC/ SE) and the National Council for Scientific and Technological Development (CNPq), under grant Nos. 794017/2013 and 444372/2024-5. F.A. acknowledges the grant RYC2021-031683-I funded by MCIN/AEI/10.13039/501100011033 and by the European Union's NextGenerationEU/PRTR. J.A.S.A. is supported by the National Natural Science Foundation of China under grant No. 12233001, by the National Key R\&D Program of China under grant No. 2024YFA1611602, by a Shanghai Natural Science Research Grant (24ZR1491200), by the ``111'' project of the Ministry of Education under grant No. B20019, and by the China Manned Space Project with No. CMS-CSST-2021-B03. J.A.S.A. also thanks the sponsorship from Yangyang Development Fund. G.L. acknowledges support from KICP/UChicago through a KICP Postdoctoral Fellowship. C.P. is supported by the Horizon Europe Marie Skłodowska-Curie Actions Doctoral Network MWGaiaDN (Grant agreement No. 101072454). R.M.S. acknowledges CNPq (Proc. 307726/2025-8). H.D.P. and N.I.F.S. acknowledge the support of Marco Antonio dos Santos, Danillo Lima de Sousa, Ulisses Manzo Castello, and Tajan Henrique de Amorim in providing access to and assistance with the computing facilities at the CPD-IAG.

\scriptsize{This study was financed in part by the Coordenação de Aperfeiçoamento de Pessoal de Nível Superior - Brasil (CAPES) - Finance Code 001.}

\scriptsize{This work made use of the computing facilities of the Centro de Processamento de Dados do IAG/USP (CPD-IAG), whose purchase was made possible by the Brazilian agency FAPESP (grants 2019/25950-8, 2017/24954-4, and 2009/54006-4). This work was partially supported by the Spanish MICIN/AEI/10.13039/501100011033 and by ‘ERDF A way of making Europe’ by the European Union through grant PID2024-157964OB-C21, and the Institute of Cosmos Sciences at the University of Barcelona (ICCUB, Unidad de Excelencia Mar\'{\i}a de Maeztu) through grant CEX2024-001451-M and the project 2021-SGR-00679 GRC funded by the Agència de Gestió d'Ajuts Universitaris i de Recerca (Generalitat de Catalunya).}

\scriptsize{Funding for the Sloan Digital Sky Survey V has been provided by the Alfred P. Sloan Foundation, the Heising-Simons Foundation, the National Science Foundation, and the Participating Institutions. SDSS acknowledges support and resources from the Center for High-Performance Computing at the University of Utah. SDSS telescopes are located at Apache Point Observatory, funded by the Astrophysical Research Consortium and operated by New Mexico State University, and at Las Campanas Observatory, operated by the Carnegie Institution for Science. The SDSS website is \url{www.sdss.org}. SDSS is managed by the Astrophysical Research Consortium for the Participating Institutions of the SDSS Collaboration, including the Carnegie Institution for Science, Chilean National Time Allocation Committee (CNTAC) ratified researchers, Caltech, the Gotham Participation Group, Harvard University, Heidelberg University, The Flatiron Institute, The Johns Hopkins University, L'Ecole polytechnique f\'{e}d\'{e}rale de Lausanne (EPFL), Leibniz-Institut f\"{u}r Astrophysik Potsdam (AIP), Max-Planck-Institut f\"{u}r Astronomie (MPIA Heidelberg), Max-Planck-Institut f\"{u}r Extraterrestrische Physik (MPE), Nanjing University, National Astronomical Observatories of China (NAOC), New Mexico State University, The Ohio State University, Pennsylvania State University, Smithsonian Astrophysical Observatory, Space Telescope Science Institute (STScI), the Stellar Astrophysics Participation Group, Universidad Nacional Aut\'{o}noma de M\'{e}xico, University of Arizona, University of Colorado Boulder, University of Illinois at Urbana-Champaign, University of Toronto, University of Utah, University of Virginia, Yale University, and Yunnan University.}

\scriptsize{This research used data obtained with the Dark Energy Spectroscopic Instrument (DESI). DESI construction and operations are managed by the Lawrence Berkeley National Laboratory. This material is based upon work supported by the U.S. Department of Energy, Office of Science, Office of High-Energy Physics, under Contract No. DE–AC02–05CH11231, and by the National Energy Research Scientific Computing Center, a DOE Office of Science User Facility under the same contract. Additional support for DESI was provided by the U.S. National Science Foundation (NSF), Division of Astronomical Sciences under Contract No. AST-0950945 to the NSF’s National Optical-Infrared Astronomy Research Laboratory; the Science and Technology Facilities Council of the United Kingdom; the Gordon and Betty Moore Foundation; the Heising-Simons Foundation; the French Alternative Energies and Atomic Energy Commission (CEA); the National Council of Humanities, Science and Technology of Mexico (CONAHCYT); the Ministry of Science and Innovation of Spain (MICINN), and by the DESI Member Institutions: \url{www.desi.lbl.gov/collaborating-institutions}. The DESI collaboration is honored to be permitted to conduct scientific research on I’oligam Du’ag (Kitt Peak), a mountain with particular significance to the Tohono O’odham Nation. Any opinions, findings, and conclusions or recommendations expressed in this material are those of the author(s) and do not necessarily reflect the views of the U.S. National Science Foundation, the U.S. Department of Energy, or any of the listed funding agencies.}

\scriptsize{Based on data obtained from the ESO Science Archive Facility with DOI: \url{https://doi.org/10.18727/archive/25}.}

\scriptsize{This work has made use of data from the European Space Agency (ESA) mission {\it Gaia} (\url{https://www.cosmos.esa.int/gaia}), processed by the {\it Gaia} Data Processing and Analysis Consortium (DPAC, \url{https://www.cosmos.esa.int/web/gaia/dpac/consortium}). Funding for the DPAC has been provided by national institutions, in particular the institutions participating in the {\it Gaia} Multilateral Agreement.}

\scriptsize{This work made use of the Fourth Data Release of the GALAH Survey \citep{GALAHDR4}. The GALAH Survey is based on data acquired through the Australian Astronomical Observatory, under programs: A/2013B/13 (The GALAH pilot survey); A/2014A/25, A/2015A/19, A2017A/18 (The GALAH survey phase 1); A2018A/18 (Open clusters with HERMES); A2019A/1 (Hierarchical star formation in Ori OB1); A2019A/15, A/2020B/23, R/2022B/5, R/2023A/4, R2023B/5 (The GALAH survey phase 2); A/2015B/19, A/2016A/22, A/2016B/10, A/2017B/16, A/2018B/15 (The HERMES-TESS program); A/2015A/3, A/2015B/1, A/2015B/19, A/2016A/22, A/2016B/12, A/2017A/14, A/2020B/14 (The HERMES K2-follow-up program); R/2022B/02 and A/2023A/09 (Combining asteroseismology and spectroscopy in K2); A/2023A/8 (Resolving the chemical fingerprints of Milky Way mergers); and A/2023B/4 (s-process variations in southern globular clusters). We acknowledge the traditional owners of the land on which the AAT stands, the Gamilaraay people, and pay our respects to elders past and present. This paper includes data that has been provided by AAO Data Central (\url{datacentral.org.au)}.}

\scriptsize{Guoshoujing Telescope (the Large Sky Area Multi-Object Fiber Spectroscopic Telescope LAMOST) is a National Major Scientific Project built by the Chinese Academy of Sciences. Funding for the project has been provided by the National Development and Reform Commission. LAMOST is operated and managed by the National Astronomical Observatories, Chinese Academy of Sciences.}

\scriptsize{Funding for RAVE has been provided by: the Australian Astronomical Observatory; the Leibniz-Institut fuer Astrophysik Potsdam (AIP); the Australian National University; the Australian Research Council; the French National Research Agency; the German Research Foundation (SPP 1177 and SFB 881); the European Research Council (ERC-StG 240271 Galactica); the Istituto Nazionale di Astrofisica at Padova; The Johns Hopkins University; the National Science Foundation of the USA (AST-0908326); the W. M. Keck foundation; the Macquarie University; the Netherlands Research School for Astronomy; the Natural Sciences and Engineering Research Council of Canada; the Slovenian Research Agency (research core funding No. P1-0188); the Swiss National Science Foundation; the Science \& Technology Facilities Council of the UK; Opticon; Strasbourg Observatory; and the Universities of Groningen, Heidelberg and Sydney. The RAVE web site is \url{https://www.rave-survey.org}.}

\scriptsize{Funding for SDSS-III has been provided by the Alfred P. Sloan Foundation, the Participating Institutions, the National Science Foundation, and the U.S. Department of Energy Office of Science. The SDSS-III website is \url{http://www.sdss3.org/}. SDSS-III is managed by the Astrophysical Research Consortium for the Participating Institutions of the SDSS-III Collaboration including the University of Arizona, the Brazilian Participation Group, Brookhaven National Laboratory, Carnegie Mellon University, University of Florida, the French Participation Group, the German Participation Group, Harvard University, the Instituto de Astrofisica de Canarias, the Michigan State/Notre Dame/JINA Participation Group, Johns Hopkins University, Lawrence Berkeley National Laboratory, Max Planck Institute for Astrophysics, Max Planck Institute for Extraterrestrial Physics, New Mexico State University, New York University, Ohio State University, Pennsylvania State University, University of Portsmouth, Princeton University, the Spanish Participation Group, University of Tokyo, University of Utah, Vanderbilt University, University of Virginia, University of Washington, and Yale University.}

\end{acknowledgments}


\appendix
\twocolumngrid

\section{Comparison between StarHorse and Bailer-Jones}
\label{appendix_comparison_SH_BJ}

\begin{figure*}[htb!]
    \centering
    \includegraphics[width=0.9\linewidth]{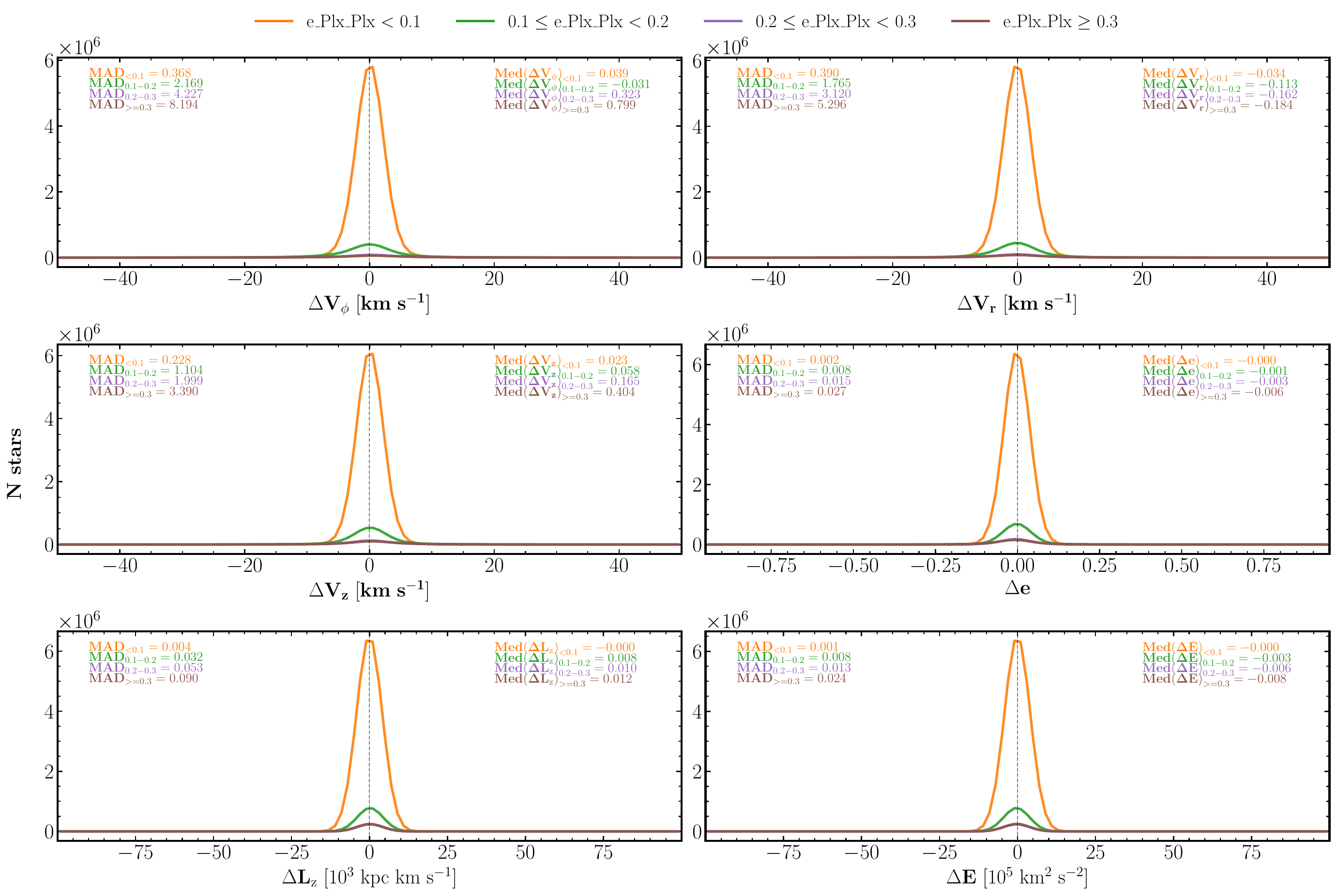}
    \caption{Distribution of the differences between SH- and BJ-derived orbital parameters $V_{\phi}$ (upper left panel), $V_{r}$ (upper right panel), $V_{z}$ (center left panel), eccentricity (center right panel), $L_{z}$ (lower left panel) and energy (lower right panel) for the \texttt{Full} sample (in the sense SH $-$ BJ). The orange, green, purple, and brown lines represent stars with  $\sigma_{\varpi}/\varpi < 0.1$, $0.1 < \sigma_{\varpi}/\varpi < 0.2$,  $0.2< \sigma_{\varpi}/\varpi < 0.3$, and $\sigma_{\varpi}/\varpi > 0.3$. The numbers in the upper-left and upper-right corners of each panel are the MAD and the median of the difference, respectively.}
    \label{fig:comp_SH_BJ}
\end{figure*}

To analyze whether the use of different distance estimates would significantly affect our orbital parameters, we compare the estimates obtained using SH and BJ distances. For illustration, Figure~\ref{fig:comp_SH_BJ} shows the distributions of the differences in $V_{\phi}$, $V_{r}$, $V_{z}$, eccentricity, $L_z$, and energy for different ranges of the parallax fractional uncertainty, $\sigma_{\varpi}/\varpi$. The comparison reveals no significant systematic offset between the two distance estimates. However, the dispersion of the differences increases with parallax uncertainty, particularly for $V_{\phi}$, $V_{r}$, $V_{z}$, and orbital eccentricity. These results demonstrate that the choice of distance estimator becomes increasingly important in the low signal-to-noise parallax regime.

\section{Orbital Parameters Dependence of Radial-Velocity Differences}
\label{appendix_param_rv}

\begin{figure}[htb!]
    \centering
    \includegraphics[width=0.99\linewidth]{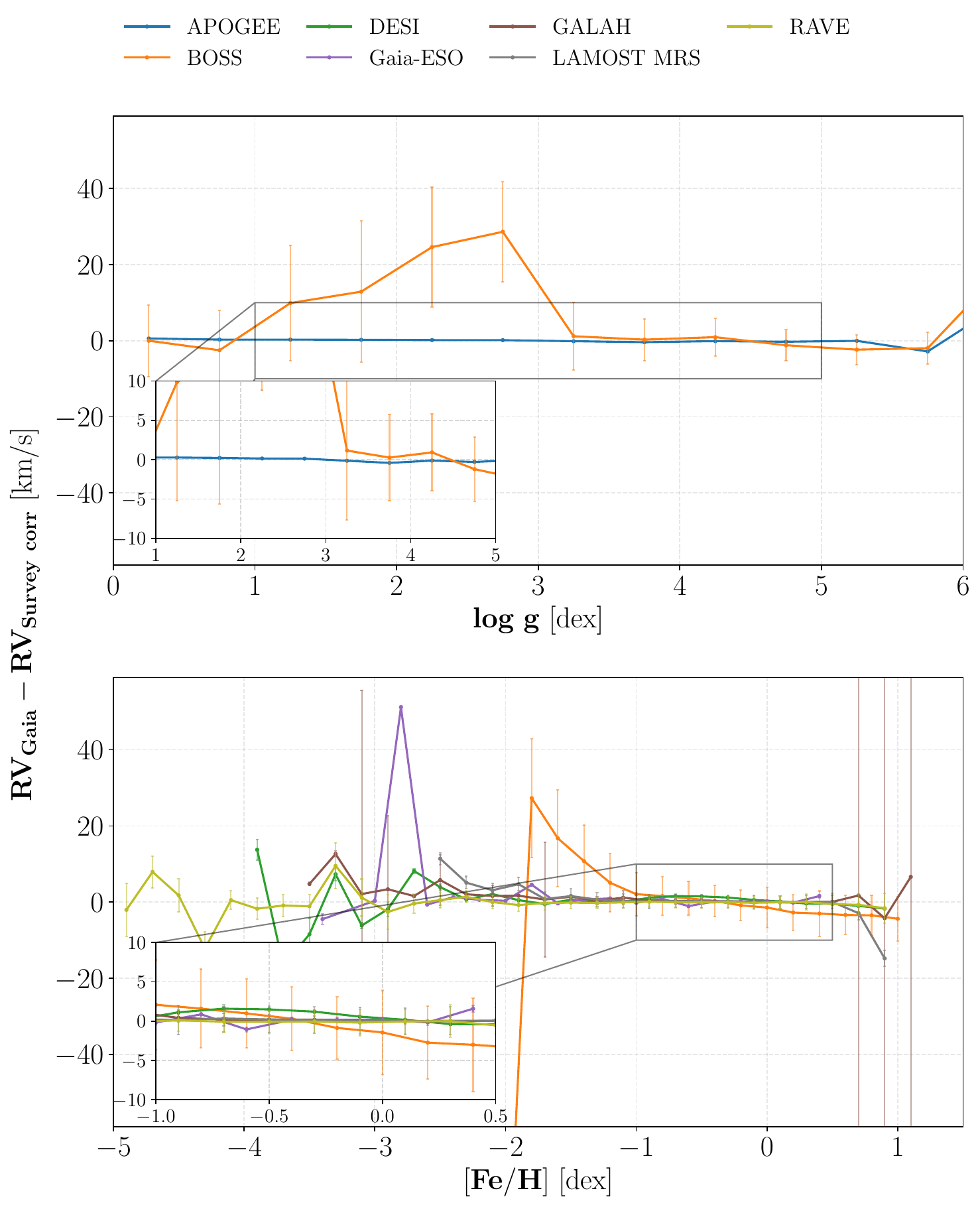}
    \caption{Systematic trends in the differences between survey-normalized radial velocities and Gaia radial velocities as a function of $\log$ g (upper panel) and metallicity (lower panel). In the upper panel, the behavior of BOSS stars (orange) is compared with that of APOGEE stars (blue). In the lower panel, results are shown for BOSS (orange), DESI (green), Gaia-ESO (purple), GALAH (brown), LAMOST MRS (gray), and RAVE (lime). The inset panels provide a closer view of the parameter ranges where no significant systematic trends are observed.}
    \label{fig:trends_rv}
\end{figure}

To further assess the consistency of the radial-velocity calibration, we examined the differences between the corrected \textit{Gaia} radial velocities and the normalized radial velocities of each survey as a function of stellar and observational parameters, including $T_{\rm eff}$, $\log g$, [Fe/H], $G_{\rm RVS}$, and SNR. Overall, the median RV differences remain approximately constant across the main, well-populated parameter ranges, with variations typically within $\sim 5$ km s$^{-1}$. Larger deviations are observed toward the extremes of some parameter distributions, particularly for metal-poor stars, faint sources, and low-SNR spectra. Examples of these extreme cases are shown in Figure~\ref{fig:trends_rv}. The most significant caveats concern BOSS objects with $\mathrm{XCSAO\_FEH} < -1.5$ and $1 < \mathrm{XCSAO\_LOGG} < 3$, corresponding to 12\% and 10\% of the final BOSS sample, respectively; DESI objects with $\mathrm{FEH} < -2.5$, representing 13\% of its final sample; Gaia-ESO objects with $\mathrm{FEH} < -2.5$, accounting for less than 1\% of its final sample; GALAH objects with $\mathrm{fe\_h} < -3$ or $\mathrm{fe\_h} > 0.7$, representing approximately 1\% of its final sample; LAMOST MRS objects with $\mathrm{feh\_cnn} > 0.5$ or $\mathrm{feh\_cnn} < -2.2$, corresponding to 18\% of its final sample; and RAVE objects with $\mathrm{m\_H\_m} < -3$, accounting for less than 1\% of its final sample. Taken together, these subsets comprise 528,690 objects for \texttt{StarHorse} distances, and 1,124,548 objects for Bailer-Jones distances, corresponding to 3.4\% of the full catalog when using SH distances and 2.6\% when using BJ distances. Larger dispersions and uncertainties generally accompany these deviations and occur in less populated regions of parameter space, particularly for some low-resolution surveys. Although residual parameter-dependent systematics may therefore affect these subsets, the results do not indicate strong trends across the main parameter ranges that would justify additional corrections. Furthermore, a test in which radial velocities were perturbed by an amount corresponding to the largest typical survey dispersion shows that the derived orbital parameters are not strongly sensitive to RV variations at this level.

\section{Number of Monte Carlo realizations}
\label{appendix_SEGUE_MC}

\begin{figure*}[htb!]
    \centering
    \includegraphics[width=0.99\linewidth]{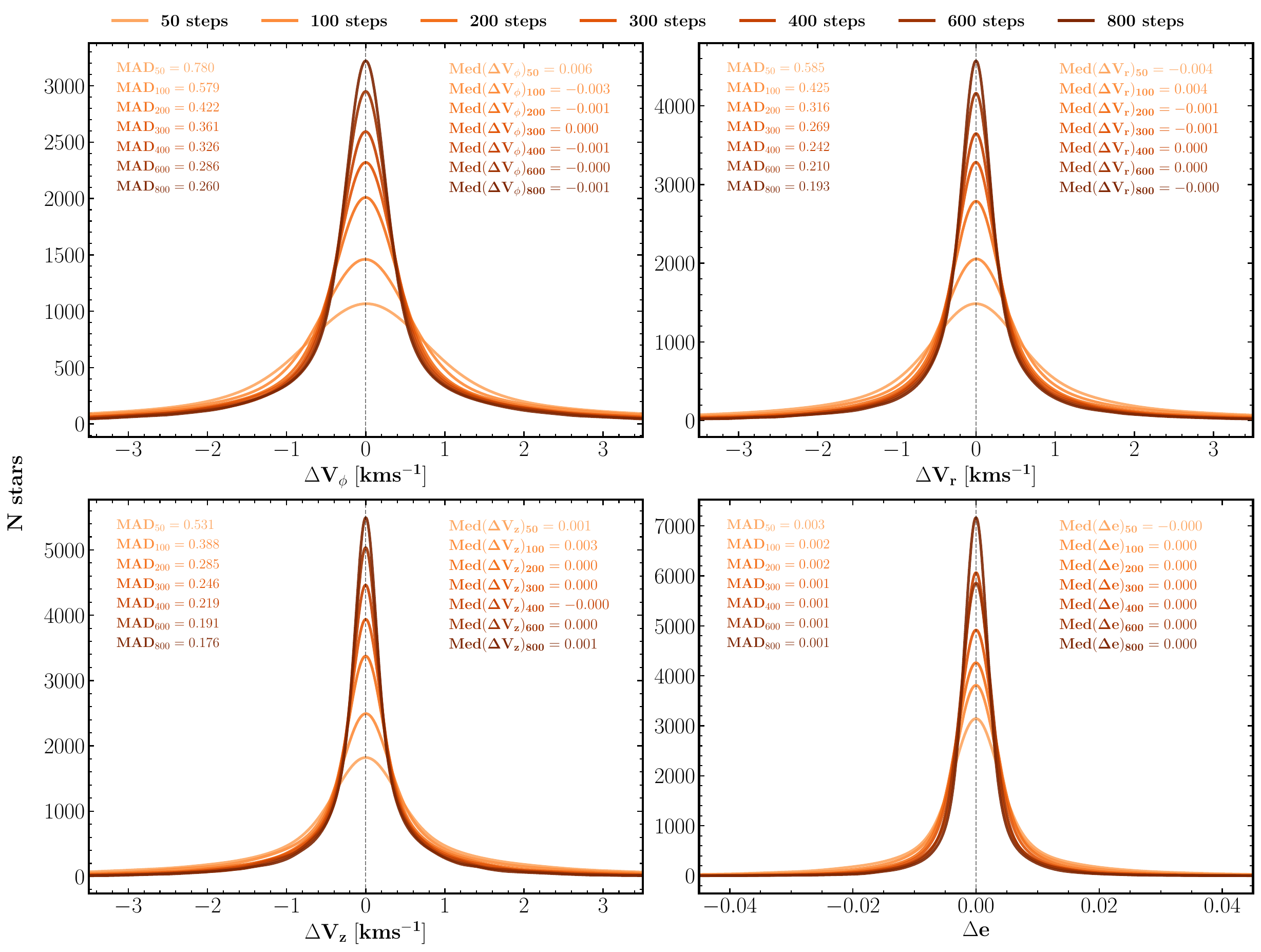}
    \caption{Distribution of the differences of $V_{\phi}$ (upper left panel), $V_r$ (upper right panel), $V_z$ (lower left panel), and $e$ (lower right panel) for a random sample of 50,000 stars from SEGUE. Each panel compares 50, 100, 200, 300, 400, 600, and 800 MC realizations, relative to the reference solution computed with 1,000 realizations. The values listed in each panel correspond to the median of each difference distribution.}
    \label{fig:test_segue}
\end{figure*}

\normalsize{In this study, we investigated the number of Monte Carlo (MC) iterations necessary to obtain statistically robust orbital parameters. For this purpose, we analyzed a sample of 50,000 random stars from the SEGUE survey, which provides sources representative of all Galactic components \citep{SEGUE1, SEGUE2}. We tested the performance of the MC algorithm with 100, 200, 300, 400, 600, 800, and 1,000 iterations, using the 1,000-iteration solution as our reference standard and comparing the others against it.

For illustrative purposes, we show in Figure \ref{fig:test_segue} the distribution of comparisons of velocities in cylindrical coordinates, $V_{\phi}$,$V_r$,  $V_z$, and eccentricity obtained in our tests. We show both MAD and median for each step compared with our 1,000-iteration standard.

Based on this comparison, we selected 100 steps as a suitable compromise between computational efficiency and result accuracy, as shown in Figure \ref{fig:test_segue}.
The orbital parameters were then determined using 100 integrations.


\section{Catalog columns}
\label{appendix_columns}

In this Appendix, we describe the columns of our catalog. The first 11 columns represent the input information, while the next 22 parameters represent the orbital parameters, obtained using \texttt{AGAMA}.  Several parameters, such as velocities, actions, energy, angular momentum, etc, are accompanied by their 16th and 84th percentiles, which we omit here for simplicity.  We marked those parameters with a star ($^{\star}$). The next three columns concern the information from the original survey. The four columns after that are the Galactic membership classification, and the last four are the flags we created. We list all available columns in Table \ref{tab:table_description}.

\renewcommand{\arraystretch}{0.85}
\begin{deluxetable*}{lp{10cm}l}
\tablecaption{Content description of the catalog. \label{tab:table_description}}
\tablehead{
\colhead{Column name} & \colhead{Description} & \colhead{Units}
}
\startdata
DR3Name                           & \textit{Gaia} DR3 source ID                                      &                 \\
RA                                & Right ascension (J2000)                                          & deg             \\
DEC                               & Declination (J2000)                                              & deg             \\
DIST                              & Photogeometric distance from \citealt{BailerJones} / Distance from \texttt{StarHorse}                 & kpc             \\
RV\_NORM \tablenotemark{$^{\dagger}$}  & Radial velocity normalized with \textit{Gaia} DR3                & km s$^{-1}$     \\
PMRA                              & Proper motion in right ascension direction                       & mas yr$^{-1}$   \\
PMDEC                             & Proper motion in declination direction                           & mas yr$^{-1}$   \\
e\_PMRA                           & Standard uncertainty of proper motion in right ascension direction     & mas yr$^{-1}$   \\
e\_PMDEC                          & Standard uncertainty of proper motion in declination direction         & mas yr$^{-1}$   \\
e\_DIST                           & Distance uncertainty of the photogeometric distance from \citealt{BailerJones} / Distance uncertainty from \texttt{StarHorse}      & kpc             \\
e\_RV \tablenotemark{$^{\dagger}$}& Radial velocity uncertainty from the original survey                   & km s$^{-1}$     \\
Survey                            & Original spectroscopic survey                                    &                 \\
Survey\_ID                        & Source ID from the original survey                               &                 \\
SNR                               & Signal-to-noise ratio from the original survey                   &                 \\
Xgal\tablenotemark{$^{\star}$}    & Galactic Cartesian position for x-axis                           & kpc             \\
Ygal\tablenotemark{$^{\star}$}    & Galactic Cartesian position for y-axis                           & kpc             \\
Zgal\tablenotemark{$^{\star}$}    & Galactic Cartesian position for z-axis                           & kpc             \\
Rperi\tablenotemark{$^{\star}$}   & Pericenter distance                                              & kpc             \\
Rapo\tablenotemark{$^{\star}$}    & Apocenter distance                                               & kpc             \\
Rmax\tablenotemark{$^{\star}$}    & Galactocentric radius                                            & kpc             \\
Zmax\tablenotemark{$^{\star}$}    & Maximum height above the Galactic plane                          & kpc             \\
Ecc\tablenotemark{$^{\star}$}     & Orbital eccentricity                                             &                 \\
Incl\tablenotemark{$^{\star}$}    & Orbital inclination                                              & deg             \\
Vx\tablenotemark{$^{\star}$}      & Velocity in Cartesian coordinates for x-axis                     & km s$^{-1}$     \\
Vy\tablenotemark{$^{\star}$}      & Velocity in Cartesian coordinates for y-axis                     & km s$^{-1}$     \\
Vr\tablenotemark{$^{\star}$}      & Radial velocity in cylindrical coordinates                       & km s$^{-1}$     \\
Vphi\tablenotemark{$^{\star}$}    & Azimuthal velocity in cylindrical coordinates                    & km s$^{-1}$     \\
Vz\tablenotemark{$^{\star}$}      & Vertical velocity in cylindrical coordinates                     & km s$^{-1}$     \\
Energy\tablenotemark{$^{\star}$}  & Total orbital energy                                             & km$^{2}$ s$^{-2}$ \\
Lx\tablenotemark{$^{\star}$}      & X-component of angular momentum in cartesian coordinates         & kpc km s$^{-1}$ \\
Ly\tablenotemark{$^{\star}$}      & Y-component of angular momentum in cartesian coordinates         & kpc km s$^{-1}$ \\
Lz\tablenotemark{$^{\star}$}      & Z-component of angular momentum in cartesian coordinates         & kpc km s$^{-1}$ \\
Ltot\tablenotemark{$^{\star}$}    & Total angular momentum                                           & kpc km s$^{-1}$ \\
Lperp\tablenotemark{$^{\star}$}   & Perpendicular angular momentum                                   & kpc km s$^{-1}$ \\
Jr\tablenotemark{$^{\star}$}      & Radial action in cylindrical coordinates                         & kpc km s$^{-1}$ \\
Jz\tablenotemark{$^{\star}$}      & Vertical action in cylindrical coordinates                       & kpc km s$^{-1}$ \\
Clean                             & Indicative flag for accurate radial velocity                    &                 \\
Unique                            & Indicative flag for best measurements for an object              &                 \\
Golden                            & Indicative flag for best measurements for an object with accurate radial velocity                     &                 \\
P\_D\_K                           & Kinematic probability of thin disk membership                    &                 \\
P\_TD\_K                          & Kinematic probability of thick disk membership                   &                 \\
P\_H\_K                           & Kinematic probability of halo membership                         &                 \\
Chemical\_membership               & Galactic component based on the chemical classification & \\
Population                        & Indicative flag for population membership                        &                 \\
Used\_mag                       & Indicative flag for adopted \textit{Gaia} magnitude in the radial velocity correction                        &                 \\ \hline
\enddata
\tablecomments{Parameters marked with \tablenotemark{$^{\star}$} are accompanied by their 16th and 84th percentiles.}
\tablecomments{\tablenotemark{$^{\dagger}$} in {\it Gaia} case those values were corrected according Section ~\ref{subsec:corrections}.}
\end{deluxetable*}
\renewcommand{\arraystretch}{1.0}

\newpage


\section{Chemical selection of Galactic components}
\label{appendix_chemical_selection}

For our chemical classification in Section \ref{subsec:chemical_class}, we used only APOGEE, GALAH, and LAMOST MRS, as they are the only ones with reliable measurements of metallicities ([Fe/H]) and alpha-element abundances  ([Mg/Fe]). We were willing to include the chemical classification for \textit{Gaia}-ESO as well. However, despite the high precision, the disk bimodality was not clearly distinguished, and we opted not to include it.

For those surveys, we apply quality cuts based on their chemical parameters. We adopted the same signal-to-noise ratio cut as before in the \texttt{Clean} sample and ensured that we use only objects with both [Fe/H] and [Mg/Fe] available. Additionally, we apply \texttt{fe\_h\_flags == 0} and \texttt{mg\_h\_flags == 0} for APOGEE, and \texttt{flag\_fe\_h == 0} and \texttt{flag\_mg\_fe == 0} for GALAH. LAMOST MRS does not provide a flag related to this. Also, for LAMOST MRS, we used the [Fe/H] from the Convolutional Neural Network (CNN) rather than the pipeline value because we only have [Mg/Fe] from the CNN and wanted to keep it consistent.

The separation between thin disk, thick disk, and halo is based on \citet{Borbolato2025}, and obeys the following equations, which better describe the separation between the populations for each survey:

\begin{itemize}
    \item APOGEE: The curve represents [Fe/H] = $-1$ for [Mg/Fe] $> 0.17$, [Mg/Fe] $= 0.18$ for the interval $-1.0 <$ [Fe/H] $< -$0.6, [Mg/Fe] = $-0.15 \times$ [Fe/H] $+ 0.09$ between $-0.6 <$ [Fe/H] $< 0.1$, [Mg/Fe] $= 0.08$ for [Fe/H] $> 0.1$, and [Mg/Fe] $= -0.556 \times$ [Fe/H] $+ 0.388$ for [Fe/H] $>-$1.
    
    \item GALAH: The separation lines are [Fe/H] = $-$1 for [Mg/Fe] $> 0.233$, [Mg/Fe] $= 0.233$ for the interval $-1.0 <$ [Fe/H] $< -0.6$, [Mg/Fe] $= -0.19 \times$ [Fe/H] $+ 0.12$ between $-0.6 <$ [Fe/H] $< 0.1$, [Mg/Fe] $= 0.101$ for [Fe/H] $> 0.1$, and [Mg/Fe] $= -0.538 \times$ [Fe/H] $+ 0.3076$ for [Fe/H] $>-1$.
    
    \item LAMOST MRS: It is separated by [Fe/H] $= -1$ for [Mg/Fe] $> 0.17$, [Mg/Fe] $= 0.18$ for the interval $-1.0 <$ [Fe/H] $< -0.6$, [Mg/Fe] = $-0.20 \times$ [Fe/H] $+ 0.06$ between $-0.6 <$ [Fe/H] $< 0.1$, [Mg/Fe]$ = 0.04$ for [Fe/H] $> 0.1$, and [Mg/Fe] $= -0.556 \times$ [Fe/H] $+ 0.388$ for [Fe/H] $> -1$.
\end{itemize}

\section{Surveys}
\label{appendix_surveys}

When individual surveys are examined separately, their distinct selection functions and spatial coverages become evident (Figures~\ref{fig:xz_xy_surveys} and\ref{fig:kinematics_surveys}). \textit{Gaia} RVS and DESI provide the broadest spatial sampling, encompassing all major Galactic components and reaching large distances from the Galactic plane, with DESI preferentially probing the inner halo. SEGUE exhibits a more fragmented and filamentary spatial distribution, reflecting its targeting of distant stars in specific sky fields. In contrast, \textit{Gaia}-ESO, GALAH, and RAVE are more concentrated toward the Galactic plane, with typically lower $|Z|$ values. APOGEE partially covers the bulge region and extends toward the Galactic anticenter, while LAMOST LRS and MRS display wide spatial distributions along the Galactic $X$ axis, tracing predominantly thick-disk populations.

By combining velocity, energy, and action spaces, it becomes possible to identify kinematic structures that are not immediately apparent from spatial distributions alone. Figure~\ref{fig:kinematics_surveys} illustrates this for APOGEE, showing the $V_R$–$V_\phi$ (left), $E$–$L_Z$ (middle), and $L_Z$–eccentricity (right) planes. The $V_R$–$V_\phi$ diagram highlights more clearly the concentration of thin-disk stars in surveys such as \textit{Gaia}-ESO, LAMOST MRS, and RAVE. Consistent with their selection functions, \textit{Gaia}-ESO, GALAH, LAMOST MRS, and RAVE are dominated by stars on prograde orbits, characteristic of thin-disk kinematics. In contrast, DESI, LAMOST LRS, and SEGUE contain a substantial fraction of stars with halo-like properties.

Although some regions of parameter space remain dominated by specific populations, a significant overlap between components is present, particularly in the \textit{Gaia} and DESI samples. Part of this dispersion likely reflects increased uncertainties in distances and velocities, especially for faint and distant stars. Nevertheless, several surveys—most notably APOGEE, \textit{Gaia}-ESO, and GALAH—exhibit clear overdensities in velocity space associated with globular clusters and stellar streams, highlighting the impact of targeted observing strategies.

\begin{figure*}[htb!]
    \centering
    \includegraphics[width=0.7\linewidth]{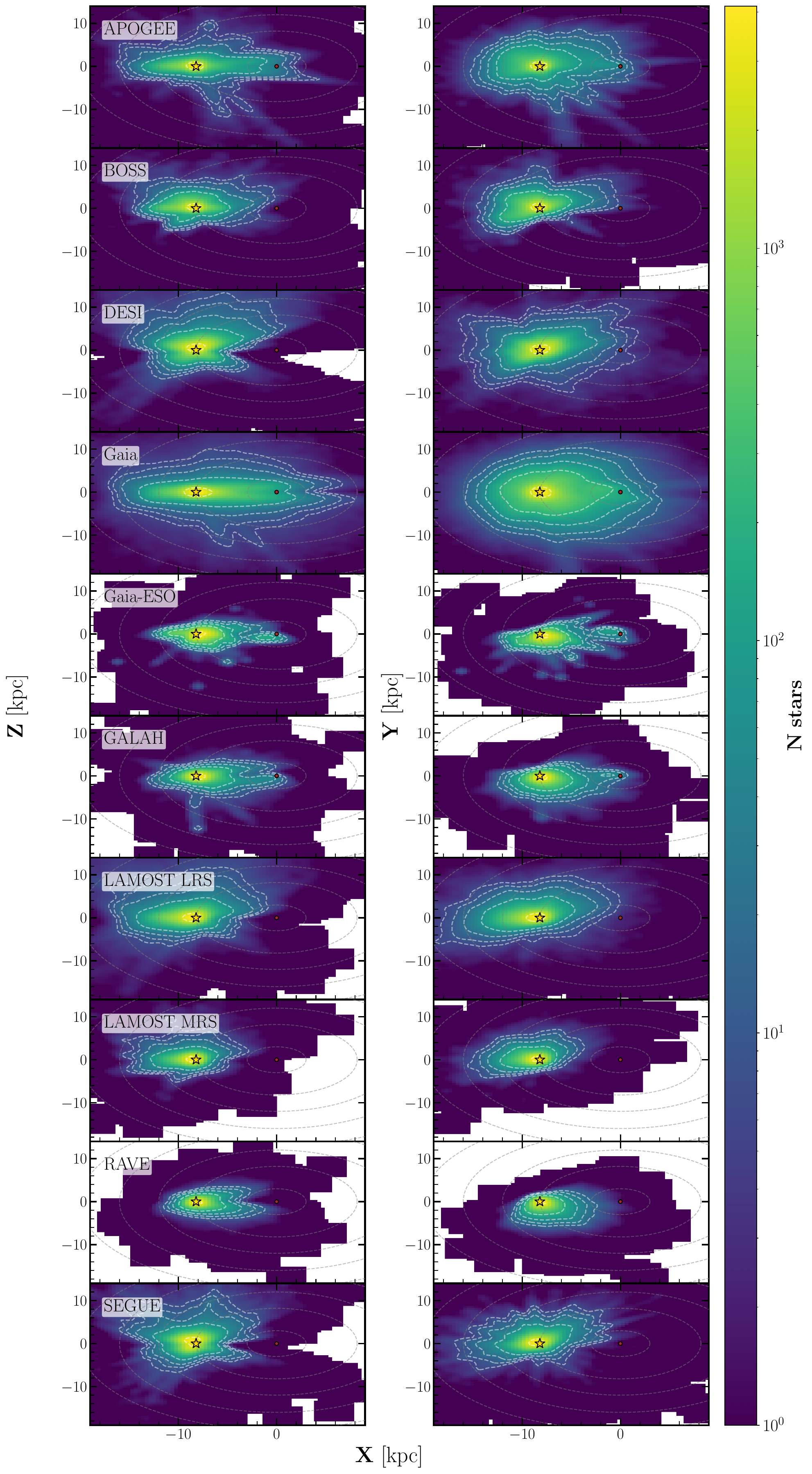}
    \caption{Position graphics, X \textit{vs.} Z (left panels) and X \textit{vs.} Y (right panels) for each spectroscopic survey used in this work, in Galactocentric coordinates. The yellow star indicates the Sun's position, and the maroon point represents the Galactic center.}
    \label{fig:xz_xy_surveys}
\end{figure*}

\begin{figure*}[htb!]
    \centering
    \includegraphics[width=0.9\linewidth]{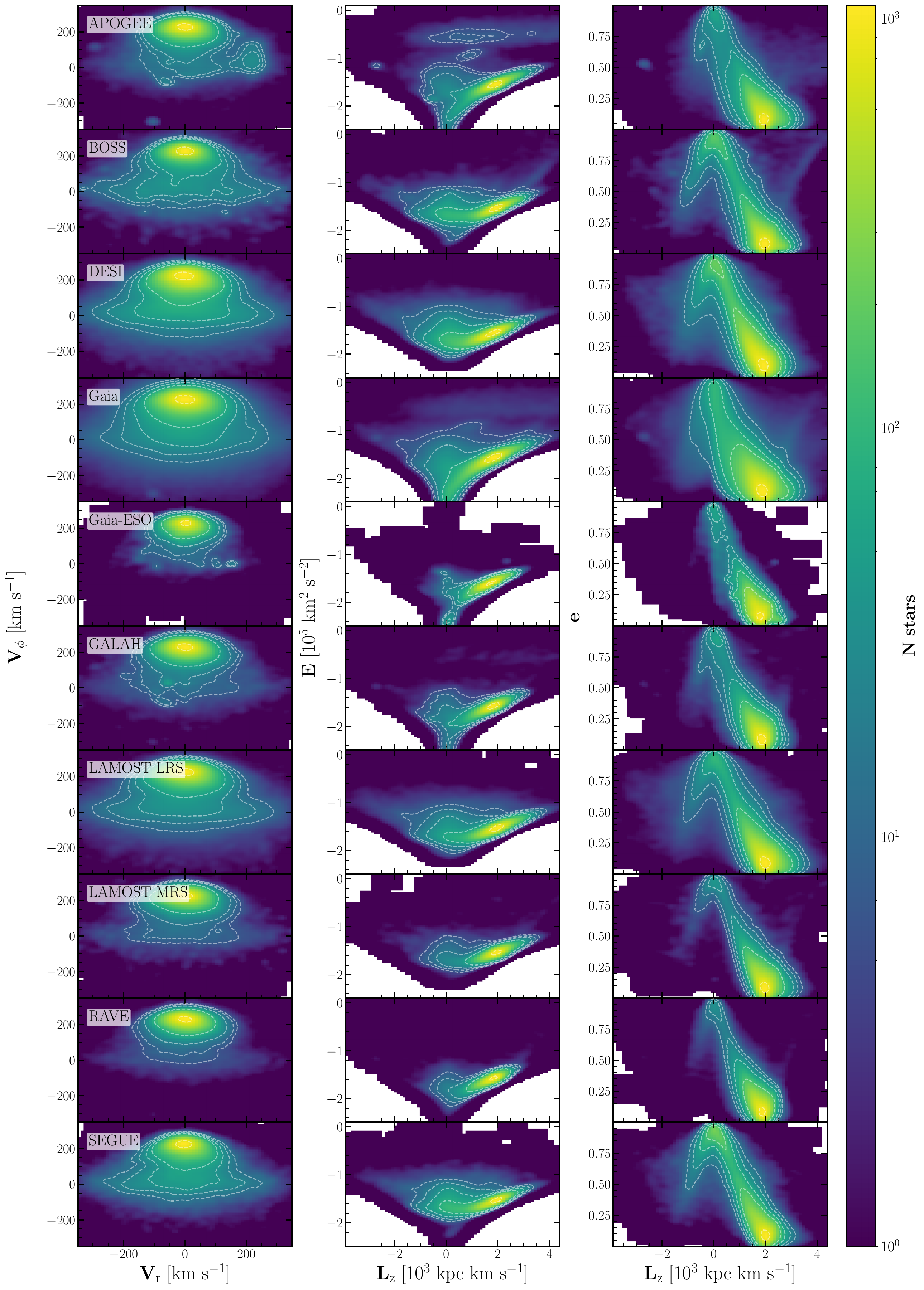}
    \caption{Orbital parameters for each spectroscopic survey used in this work, one by line, in Galactocentric coordinates. Kinematic graphics, V$_R$ \textit{vs.} V$_{\phi}$ (left panels), L$_Z$ \textit{vs.} Energy (center panels), and L$_Z$ \textit{vs.} $e$ (right panels).}
    \label{fig:kinematics_surveys}
\end{figure*}


\FloatBarrier

\bibliography{biblio2026}{}
\bibliographystyle{aasjournalv7.1}

\end{document}